\documentclass[a4paper]{article}
\usepackage{graphicx,fix-cm}
\usepackage{color}
\usepackage{amssymb,amsmath,amsfonts,bm}
\usepackage{natbib}
\usepackage[]{subfig}
\usepackage{lineno}

\input{preamble}

\graphicspath{{figures/}}

\begin{document}

\title{Clustering using skewed multivariate heavy tailed distributions with flexible tail behaviour}
\author{Darren Wraith, Florence Forbes \\ INRIA, Laboratoire Jean Kuntzman, Mistis team\\
655 avenue de l'Europe, Montbonnot\\ 38334 Saint-Ismier Cedex, France }
\date{February 23, 2014}


\maketitle

\begin{abstract}
The family of location and scale mixtures of Gaussians has the
ability to generate a number of flexible distributional forms. It
nests as particular cases several important asymmetric
distributions like the Generalised Hyperbolic distribution. The
Generalised Hyperbolic distribution in turn nests many other well
known distributions such as the Normal Inverse Gaussian (NIG)
whose practical relevance has been widely documented in the
literature. In a multivariate setting, we propose to extend the
standard location and scale mixture concept into a so called
multiple scaled framework which has the advantage of allowing
different tail and skewness behaviours in each dimension of the
variable space with arbitrary correlation between dimensions.  
Estimation of the parameters is provided via an EM algorithm with
a particular focus on NIG distributions. Inference is then
extended to cover the case of mixtures of such multiple scaled
distributions for application to clustering. Assessments on
simulated and real data confirm the gain in degrees of freedom and
flexibility in modelling data of varying tail behaviour and
directional shape.
\end{abstract}

\section{Introduction}

A popular approach to identify groups or clusters within data is
via a parametric finite mixture model \citep{fruwirth06}. While
the vast majority of work on such mixtures has been based on
Gaussian mixture models \citep[see {\it e.g.}][]{FraleyRaftery2002}.
in many applications the tails of Gaussian distributions are
shorter than appropriate and the Gaussian shape is not suitable for
highly asymmetric data.
  A natural extension to the Gaussian case is to consider families of distributions which can be represented as
   \textit{location \textbf{and} scale Gaussian mixtures} of the form,
 \begin{equation}
  p(\yv ;\mub, \Sigmab, \betab, \thetab) = \displaystyle {\int_{0}^{\infty}}
\mathcal{N}_M(\yv;\mub + w \betab\Sigmab, w \Sigmab )\;
f_W(w;\thetab) \; \text{d}w ,
\label{eqn:locscalemix}
\end{equation}
where $\mathcal{N}_M(\; . \; ;\mub + w\betab \Sigmab,w \Sigmab)$
denotes the $M$-dimensional Gaussian distribution with mean $\mub
+ w \betab\Sigmab$ and covariance $w \Sigmab$ and $f_W$ is the
probability distribution of a univariate positive variable $W$
referred to hereafter as the weight variable. The parameter
$\betab$ is an additional $M$-dimensional vector parameter for
skewness.
When $\betab=0$ and $W^{-1}$ follows a Gamma distribution ${\cal
G}(\nu/2, \nu/2)$\footnote{The Gamma probability density function
is ${\cal G}(w ; \alpha, \beta)= \beta^\alpha \Gamma(\alpha)^{-1}
w^{\alpha-1} \exp(-\beta w)$} {\it i.e.} $f_W$ is an Inverse Gamma
distribution $inv{\cal G}(\nu/2, \nu/2)$ where $\nu$ denotes the
degrees of freedom, we recover the well known multivariate
$t$-distribution \citep{KotzNadarajah2004}. The weight variable
$W$ in this case effectively acts to govern the tail behaviour of
the distributional form from light tails ($\nu \rightarrow
\infty$) to heavy tails ($\nu \rightarrow 0$) depending on the
value of $\nu$ (a form of robust tuning parameter).

In the more general case of, for example, allowing $\betab \neq 0$
and $f_W$ being a Generalised Inverse Gaussian (GIG) distribution,
we recover the family of Generalised Hyperbolic (GH) distributions
\citep{barndorffnielsen97} which is able to represent a
particularly large number of distributional forms. The GIG
distribution depends on three parameters and is given by
\begin{eqnarray}
f_W(w;\lambda,\gamma,\delta) & = & \mathcal{GIG}(w; \lambda,\gamma,\delta) \nonumber \\
& = &
\left(\frac{\gamma}{\delta}\right)^{\lambda}\frac{w^{\lambda-1}}{2K_{\lambda}(\delta\gamma)}
\text{exp}(-\frac{1}{2}(\delta^2/ w+\gamma^2 w)) \; ,
\label{eqn:gig}
\end{eqnarray}
where $K_{r}(x)$ is
 the  modified Bessel function of the third kind of order $r$ evaluated at $x$ \footnote{The modified Bessel function (see Appendix in \cite{Jorgensen1982}) is
 $K_r(x)= 1/2 \int_{0}^{\infty}y^{r-1} \exp(-\frac{1}{2} x (y+y^{-1})) \; dy $}.
Depending on the parameter choice for the GIG, special cases of
the GH family include: the multivariate GH distribution with
hyperbolic margins ($\lambda=1$) \citep{schmidtetal06}; the Normal Inverse Gaussian
 ($\lambda=-1/2$) distribution \citep{Barndorff1982}; the multivariate hyperbolic
($\lambda=\frac{M+1}{2}$) distribution \citep{Barndorff1977}; the hyperboloid
 ($\lambda=0$) distribution \citep{Jensen1981}; the hyperbolic skew-t 
($\lambda=-\nu, \gamma=0$) distribution \citep{aasetal05}; and the Normal Gamma
 ($\lambda > 0, \mub=0, \delta=0$) distribution \citep{Griffin2010} amongst others.
For applied problems, the most popular of these forms appears to
be the Normal Inverse Gaussian (NIG) distribution
\citep{Barndorff1982,protassov04,KarlisSantourian2009}. It has been used
extensively in financial applications, see
\citet{protassov04,barndorffnielsen97} or \citet{aasetal05,aashaff06} and references therein, but also in geoscience and signal processing \citep{Gjerde2011,Oigard2004}. Another popular distributional
form allowing for skewness and heavy or light tails includes
different forms of the multivariate skew-$t$ like the proposals of
\cite{sahuetal03,leemclachlan12,lin10} for skew-$t$ distributions
where skewness and covariance are separated and
\cite{azzalinietal96,bassoetal10,pyneetal09} for other
formulations that do not share this separation property. Most of
these distributional forms are also able to be represented as
\textit{location and scale Gaussian mixtures}.

Although the above approaches provide for great flexibility in
modelling data of highly asymmetric and heavy tailed form,
 they assume $f_W$ to be a univariate distribution and hence each
dimension is governed by the same amount of tailweight. There have
been various approaches to address this issue in the statistics
literature for both symmetric and asymmetric distributional forms.
In his work, \citet{Jones2002} proposes a dependent bivariate
$t$-distribution with marginals of different degrees of freedom
but the tractability of the extension to the multivariate case is
unclear. Other distributions have been presented in chapters 4 and 5 of
\citet{KotzNadarajah2004} but their formulations tend to be
appreciably more complicated, often already in the expression of
their probability density function. Increasingly, there has been
much research on copula approaches to account for flexible
distributional forms but the choice as to which one to use in this
case and the applicability to (even) moderate dimensions is also
not clear \citep{Daul2003,Giordani2008,DemartaMcNeil2005}.
In general the papers take various approaches whose relationships
have been characterized in the bivariate case by
\citet{ShawLee2008}. However, most of these approaches suffer
either from the non-existence of a closed-form pdf or from a
difficult generalization to more than two dimensions. An
alternative approach \citep{schmidtetal06}, which takes advantage
of the property that Generalised Hyperbolic distributions are
closed under affine-linear transformations, derives independent GH
marginals but estimation of parameters appears to be restricted to
density estimation, and not formally generalisable to estimation
settings for a broad range of applications ({\it e.g.} clustering,
regression, {\it etc.}).   A more general approach outside of the
GH distribution setting is outlined in
\citep{ferreirasteel07,ferreirasteel07b} with a particular focus
on regression models using a Bayesian framework.

In this paper, we show that the location and scale mixture representation can be further
 explored and propose a so-called {\it multiple scaled} framework that is considerably simpler than those previously proposed with distributions
 exhibiting interesting properties.  
The approach builds upon, and develops further, previous work on scale mixture of Gaussians \citep{forbeswraith13} where the focus was on {\it symmetric} multiple scaled  heavy tailed distributions. In this paper, we consider
the more general case of including location in addition to scale in the  multipled scaled framework. This generalisation provides a much wider variety of distributional forms, allowing different tail and skewness behavior in each dimension of the variable space with arbitrary correlation between dimensions.
The key elements of the approach are similar to that in \citet{forbeswraith13}. The introduction of multidimensional weights and a decomposition of the matrix $\Sigmab$ in \eqref{eqn:locscalemix} is used to
facilitate estimation and also allows for arbitrary correlation between dimensions.  This principle was illustrated in the Supplementary Materials (one-page Appendix B) of \citet{forbeswraith13} with the example of the NIG distribution. However, no details were given on the properties, estimation and application of these new location and scale representations. The content of this paper is therefore entirely new.  
Using the Generalised Hyperbolic  distribution as an example, we present the more general case of multiple scaled Generalised Hyperbolic
distributions for which we provide a number of properties in Sections \ref{sec:multnig} to \ref{sec:pro}. 

The paper is outlined as follows. In Section~\ref{sec:method},
further details of the GH distribution and the particular case of
the NIG distribution are briefly outlined, followed by details of
the proposed new family of multiple scaled GH (and NIG)
distributions. In Section~\ref{sec:ML}, we outline an approach for
maximum likelihood estimation of the parameters for the multiple
scaled NIG distribution via the EM algorithm. In
Section~\ref{sec:appl} we explore the performance of the approach
on several simulated and real data sets in the context of
clustering. Section~\ref{sec:disc} concludes with a discussion and
areas for further research.

\section{Multiple scaled Generalised Hyperbolic distributions}\label{sec:method}

In this section we outline further details of the standard (single
weight) multivariate GH distribution (Sect. \ref{sec:nig}) and
then the proposed multiple scaled GH distribution (Sect.
\ref{sec:multnig} to \ref{sec:pro}). As the NIG distribution
appears to be the most popular case of the GH family in
applications we also outline further details of this distribution
and its multiple scaled form which will be used in
Section~\ref{sec:appl} to assess the performance on simulated and
real datasets.

\subsection{Multivariate Generalised Hyperbolic distribution}\label{sec:nig}

As mentioned previously the Generalised Hyperbolic distribution
can be represented in terms of a \textit{location \textbf{and}
scale Gaussian mixture}. In the statistics literature, the
representation is also often referred to as a normal mean-variance
mixture. Using notation equivalent to that of
\cite{barndorffnielsen97} Section 7 and \cite{protassov04}, the
multivariate GH density takes the following form

\begin{eqnarray}
p(\yv; \mub,\Sigmab,\betab, \lambda, \gamma, \delta) &=& {\cal GH}(\yv; \mub,\Sigmab, \betab, \lambda, \gamma, \delta) \nonumber \\
& = & \int_{0}^{\infty}\mathcal{N}_{M}(\yv; \mub+w\Sigmab\betab,w\Sigmab) \; \mathcal{GIG}(w;\lambda,\gamma,\delta) dw  \nonumber \\
 &=& (2\pi)^{-M/2} |\Sigmab|^{-1/2} \left(\frac{\gamma}{\delta}\right)^{\lambda}\bigg(\frac{q(\yv)}{\alpha}\bigg)^{\lambda-\frac{M}{2}}K_{\lambda-\frac{M}{2}}(q(\yv)\alpha) \nonumber \\
 & & \times \hspace{2mm}
 (K_{\lambda}(\delta\gamma))^{-1}\textup{exp}(\betab^T(\yv-\mub))
 \label{eqn:gh}
\end{eqnarray}
where
 $|\Sigmab|$ denotes the determinant of $\Sigmab$, $\delta>0$,
 and $q(\yv)$ and $\alpha$ are given by
 \begin{eqnarray}
 q(\yv)^2 &=& \delta^2 + (\yv-\mub)^T\Sigmab^{-1}(\yv-\mub) \;, \label{def:q}\\
  \gamma^2 &=& \alpha^2-\betab^T\Sigmab\betab\geqq0
  \label{def:alpha}\; .
  \end{eqnarray}

 The parameters $\betab$ and $\mub$ are column vectors of length
$M$ ($M \times 1$ vector).


An alternative (hierarchical) representation of the multivariate
GH distribution  (which is useful for simulation) can be seen as,
\begin{eqnarray}
\Yb|W=w & \sim & \mathcal{N}_M(\mub+w\Sigmab\betab,w\Sigmab) \nonumber \\
W & \sim & \mathcal{GIG}(\lambda,\gamma,\delta)
\label{eqn:gigstandard}
\end{eqnarray}

By setting $\lambda=-1/2$ in the GIG distribution we recover the
Inverse Gaussian (IG) distribution,
\begin{eqnarray}
f_W(w;\gamma,\delta) & = & \mathcal{IG}(w;\gamma,\delta) \\
& = & \frac{\delta}{w^{3/2}\sqrt{2\pi}} \;
\text{exp}(\delta\gamma) \; \text{exp}(-\frac{1}{2}(\delta^2/
w+\gamma^2 w))
\end{eqnarray}
which (when used as the mixing distribution) leads to the  NIG
distribution
\begin{eqnarray*}
p(\yv;\mub,\Sigmab,\betab,\gamma,\delta) &=& {\cal NIG}(\yv;\mub,\Sigmab,\betab, \gamma, \delta) \nonumber\\
&=& \int_{0}^{\infty}\mathcal{N}_{M}(\yv; \mub+w\Sigmab\betab,w\Sigmab) \; \mathcal{IG}(w;\gamma,\delta) dw  \\
& = & \frac{\delta}{2^{\frac{M-1}{2}}}
\text{exp}(\delta\gamma+(\yv-\mub)^{T}\betab)\bigg(\frac{\alpha}{\pi
q(\yv)}\bigg)^{\frac{M+1}{2}} K_{\frac{M+1}{2}}(\alpha q(\yv))
\end{eqnarray*}
where $\alpha$ and $q$ are defined as in
definitions~\eqref{def:alpha} and \eqref{def:q}.

Using the parameterisation of \citep{barndorffnielsen97}, an
identification problem arises as the distributions
 ${\cal GH}(\mub,\Sigmab,\betab,\lambda,\gamma, \delta)$ and
${\cal GH}(\mub,k^2 \Sigmab,\betab, \lambda,k \gamma, \delta/k)$
are identical for any $k > 0$.  For the estimation of parameters,
this problem can be solved by constraining the determinant of
$\Sigmab$ to be 1.

\subsection{Multiple Scaled Generalised Hyperbolic distribution (MSGH)}\label{sec:multnig}

As mentioned in the Introduction, most of the work on multivariate \textit{location and scale mixture of
Gaussians} has focused on studying different choices for the weight
distribution $f_W$ \citep[see {\it e.g.}][]{EltoftKimLee2006}.
Surprisingly, little work to our knowledge has focused on the
dimension of the weight variable $W$ which in most cases has been
considered as univariate. The difficulty in considering multiple
weights is the interpretation of such a multidimensional case. The
extension we propose consists then of introducing the
parameterization of the scale matrix into $ \Sigmab = \Db \Ab
\Db^{T},$ where $\Db$ is the matrix of eigenvectors of $\Sigmab$
and $\Ab$ is a diagonal matrix with the corresponding eigenvalues
of $\Sigmab$. The matrix $\Db$ determines the orientation of the
Gaussian and $\Ab$ its shape. Such a parameterization has the
advantage to allow an intuitive incorporation of the multiple
weight parameters. We propose to set the scaled covariance in
(\ref{eqn:locscalemix}) to
 $\Db \Deltab_{\wv} \Ab \Db^T \;$,
where $\Deltab_\wv=\mbox{diag}(w_{1}, \ldots, w_{M})$ is
the $ M \times M$ diagonal matrix whose diagonal components are
the weights $\{w_{1}, \ldots, w_{M}\}$.
 The generalization we propose is
therefore to define

\begin{align}
p(\yv ; \mub,\Db,\Ab,\betab, \thetab) = & {\int_{0}^{\infty}} \ldots
{\int_{0}^{\infty}} \mathcal{N}_M(\yv;\mub + \Db \Deltab_\wv \Ab
\Db^T \betab, \Db \Deltab_\wv \Ab \Db^T)   \notag \\ 
 & \times f_\wv(w_1\ldots w_M ;
\thetab) \; \text{d}w_1 \ldots d w_M \;,
\label{eqn:multnig}
\end{align}

where $\Deltab_\wv=\mbox{diag}(w_1, \ldots w_M)$, and the weights
are assumed to be independent {\it i.e.} $f_\wv(w_1\ldots,w_M;
\thetab) = f_{W_1}(w_1 ; \thetab_1) \ldots
f_{W_M}(w_M;\thetab_M)$.
Equation~\eqref{eqn:multnig} can be equivalently written as

\begin{align}
p(\yv ; \mub, \Db,\Ab, \betab, \thetab) = & \prod\limits_{m=1}^{M}
{\int_{0}^{\infty}} \mathcal{N}_1([\Db^T(\yv-\mub)]_m ; w_m A_m
[\Db^T\betab]_m, w_m A_m) \notag \\ 
 & \times f_{W_m}(w_m) \; \text{d}w_m \; 
\end{align}

where $[\Db^T(\yv-\mub)]_{m}$ denotes the
$m$th component of vector $\Db^T(\yv-\mub)$ and $A_m$ the $m$th
diagonal element  of the diagonal matrix $\Ab$ (or equivalently
the $m$th eigenvalue of $\Sigmab$).

If we set $f_{W_m}(w_m)$ to a GIG distribution ${\cal GIG}(w_m ;
\lambda_m,\gamma_m, \delta_m)$, it follows that our generalization (MSGH)
of the multivariate GH distribution with $\lambdab=[\lambda_1,
\ldots, \lambda_M]^T$ ,$\gammab= [\gamma_1, \ldots, \gamma_M]^T$
and $\deltab= [\delta_1, \ldots, \delta_M]^T$ as $M$-dimensional
vectors is:
\begin{eqnarray}
& & {\cal MSGH}(\yv ; \mub, \Db,\Ab, \betab, \lambdab, \gammab,
\deltab) \nonumber \\ 
& = & (2\pi)^{-M/2} \; \prod\limits_{m=1}^{M}\!
|A_m|^{-1/2} \;
\left(\frac{\gamma_m}{\delta_m}\right)^{\lambda_m}\bigg(\frac{q_m(\yv)}{\alpha_m}\bigg)^{\lambda_m-1/2}
\times \nonumber \\ 
& & K_{\lambda_m-1/2}(q_m(\yv) \alpha_m )(K_{\lambda_m}(\delta_m\gamma_m))^{-1}\exp([\Db^T(\yv-\mub)]_m \; [\Db^T\betab]_m)
\label{eqn:mgh}
\end{eqnarray}

  with $\alpha^2_m= \gamma_m^2 + A_m [\Db^T\betab]_m^2$ and
$q_m(\yv)^2= \delta_m^2+  A_m^{-1}[\Db^T(\yv-\mub)]_m^2\; .$

Alternatively, with $\wv=[w_1, \ldots, w_M]^T$ we can define it as
\begin{eqnarray}
\Yb|\Wb=\wv & \sim & \mathcal{N}_M(\mub+  D\Deltab_{\wv} \Ab \Db^{T}\betab,\Db\Deltab_{\wv}\Ab \Db^{T})  \nonumber \\
\Wb & \sim & {\cal GIG}(\lambda_1,\gamma_1,\delta_1) \otimes \dots
\otimes {\cal GIG}(\lambda_M,\gamma_M,\delta_M),
\label{eqn:xwmulti}
\end{eqnarray}

where notation $\otimes$ means that the $\Wb$ components are
independent. If we set $f_{W_m}(w_m)$ to an Inverse Gaussian
distribution ${\cal IG}(w_m ; \gamma_m, \delta_m)$, it follows
that our generalization (MSNIG) of the multivariate NIG distribution with
$\gammab$ and $\deltab$ as $M$-dimensional vectors is: 
\begin{eqnarray}
 {\cal MSNIG}(\yv ; \mub, \Db,\Ab, \betab,
\gammab, \deltab)\! & = & \!\prod\limits_{m=1}^{M}\!\delta_m
\exp(\delta_m \gamma_m + [\Db^T(\yv-\mub)]_m \; [\Db^T\betab]_m) \nonumber \\ 
& & \times \displaystyle \frac{\alpha_m}{\pi q_m} K_1(\alpha_m q_m(\yv)) 
\label{eqn:mnig}
\end{eqnarray}
  with $\alpha_m^2= \gamma_m^2 + A_m [\Db^T\betab]_m^2$,
$q_m(\yv)^2= \delta_m^2+  A_m^{-1}[\Db^T(\yv-\mub)]_m^2\;$ and
$K_{1}$ is the modified Bessel function of order 1.

To simulate from the MSGH distribution, it is possible to use~\eqref{eqn:xwmulti} or
\begin{equation}
\Yb = \mub + \Db\Deltab_{\wv}\Ab \Db^{T}\betab +
\Db\Ab^{1/2}[X_1\sqrt{W_1},\dots,X_M\sqrt{W_M}]^{T}
\label{eqn:multnigsim}
\end{equation}
where $\Xb \sim \mathcal{N}(0,\Ib_M)$ and $W_m \sim
\mathcal{GIG}(\lambda_m,\gamma_m,\delta_m)$ (for $m=1,\dots,M$).

It is interesting to note that the multiple scaled GH distribution
allows potentially each dimension to follow a particular case of
the GH distribution family. For example, in a bivariate setting
$\Yv=[Y_1, Y_2]^T$, the variate $Y_1$ could follow a hyperboloid
distribution ($\lambda_1$=0) and $Y_2$ a NIG distribution
($\lambda_2=-1/2$).  In the case of assuming $\lambda_m$ to be
fixed, model choice criteria such as the Bayesian Information
Criterion (BIC) could be used to discriminate between different
model families.

\subsection{Identifiability issues}
\label{sec:ident}

 In contrast to the standard multivariate GH
distribution, constraining the determinant of $\Ab$ to be 1 is not
enough to ensure identifiability in the MSGH case. Indeed,
assuming the determinant $|\Ab|=1$, if we set $\Ab', \deltab',
\gammab'$ so that $A_m'= k_m^2 A_m, \delta'_m= \delta_m/k_m$ and
$\gamma'_m=k_m \gamma_m$, for all values $k_1\ldots k_m$
satisfying $\prod_{m=1}^M k_m^2 =1$, it follows that the
determinant $|\Ab'|=1$ and that  the ${\cal MSGH}(\yv, \mub, \Db, \Ab',
\betab, \lambdab, \gammab', \deltab')$ and ${\cal MSGH}(\yv, \mub,
\Db, \Ab, \betab, \lambdab, \gammab, \deltab)$ expressions are
equal. Identifiability can be guaranteed by adding that all
$\delta_m$'s (or equivalently all $\gamma_m$'s) are equal. In
practice, we will therefore assume that for all $m=1\ldots M$,
$\delta_m=\delta$.

\subsection{Some properties of the multiple scaled GH distributions}

\label{sec:pro}

\begin{figure}[htpb]
 \centering
\begin{tabular}{ccc}
\hspace{-5mm} \includegraphics[scale=0.25]{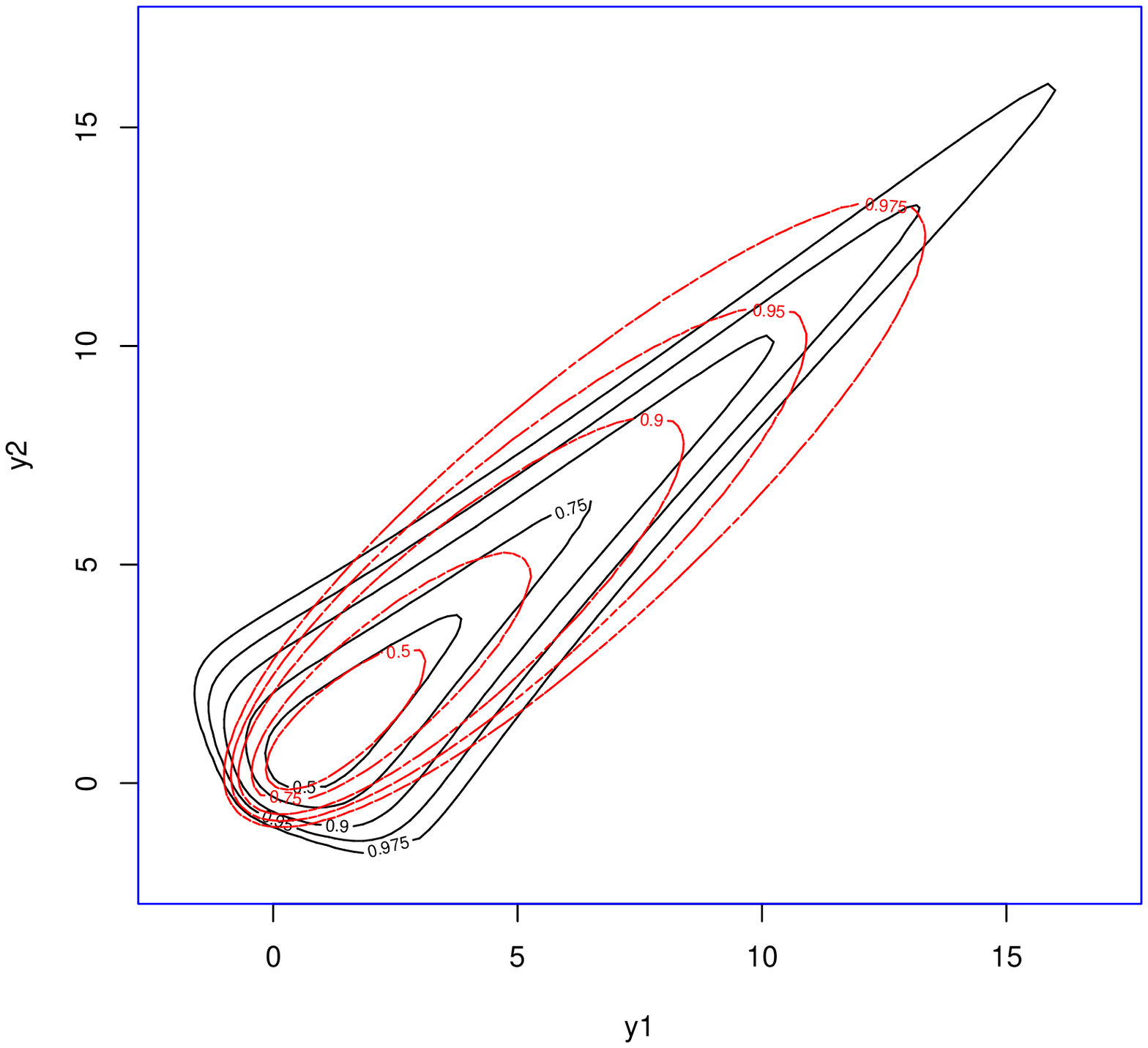} &
\hspace{-10mm} \includegraphics[scale=0.25]{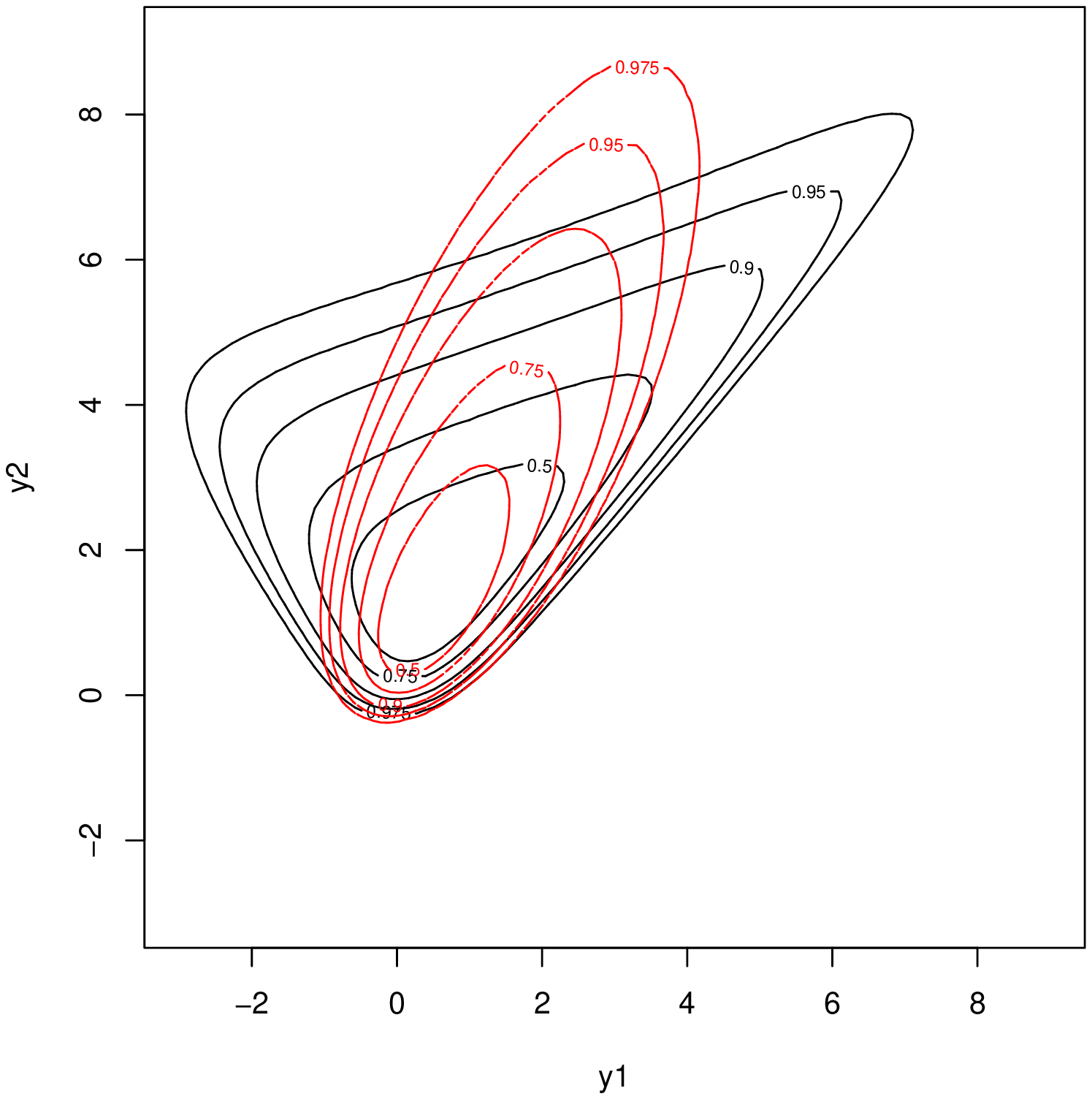} &
\hspace{-10mm} \includegraphics[scale=0.25]{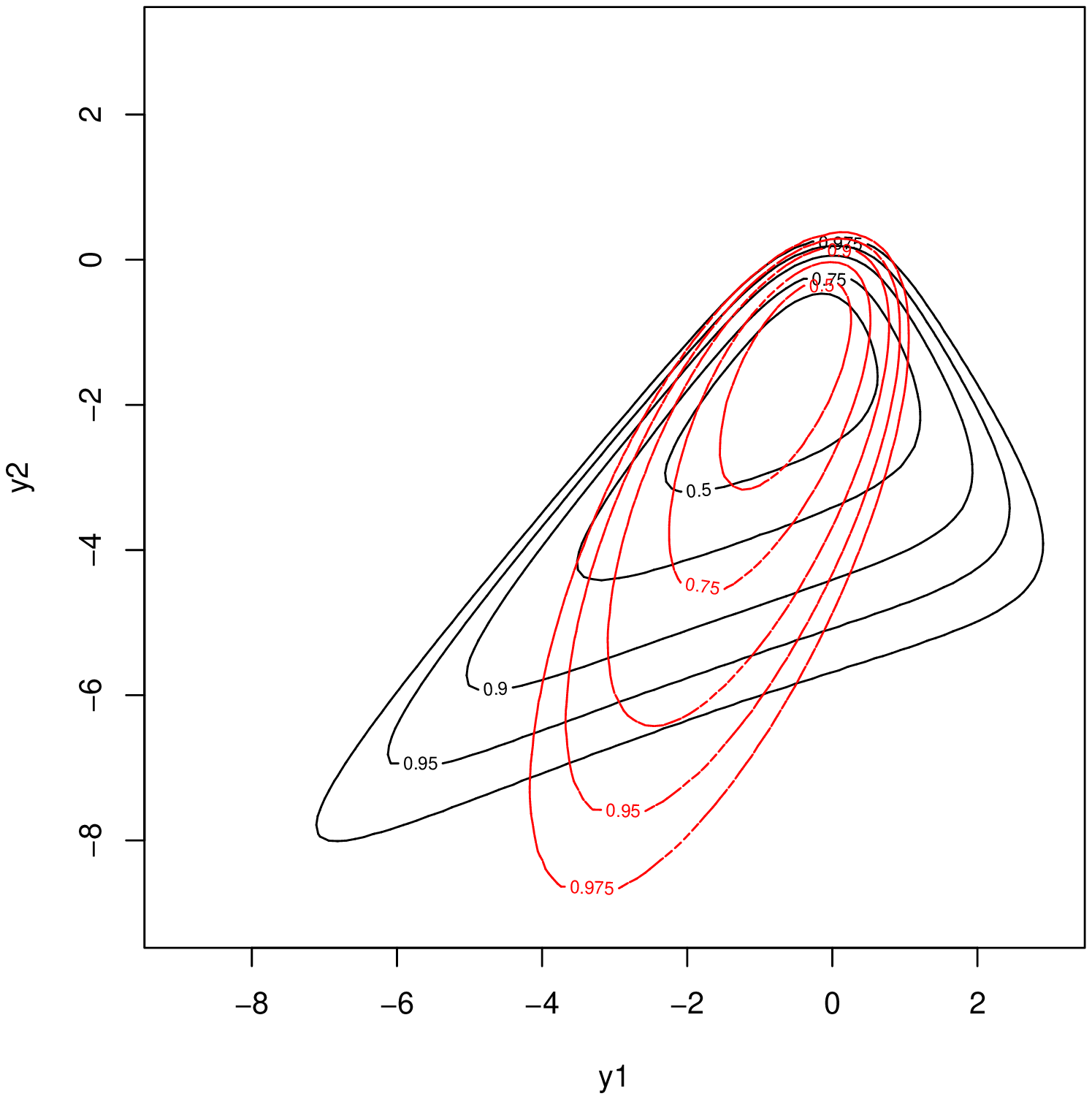} \\
(a) & (b) & (c)  \\
\hspace{-5mm} \includegraphics[scale=0.25]{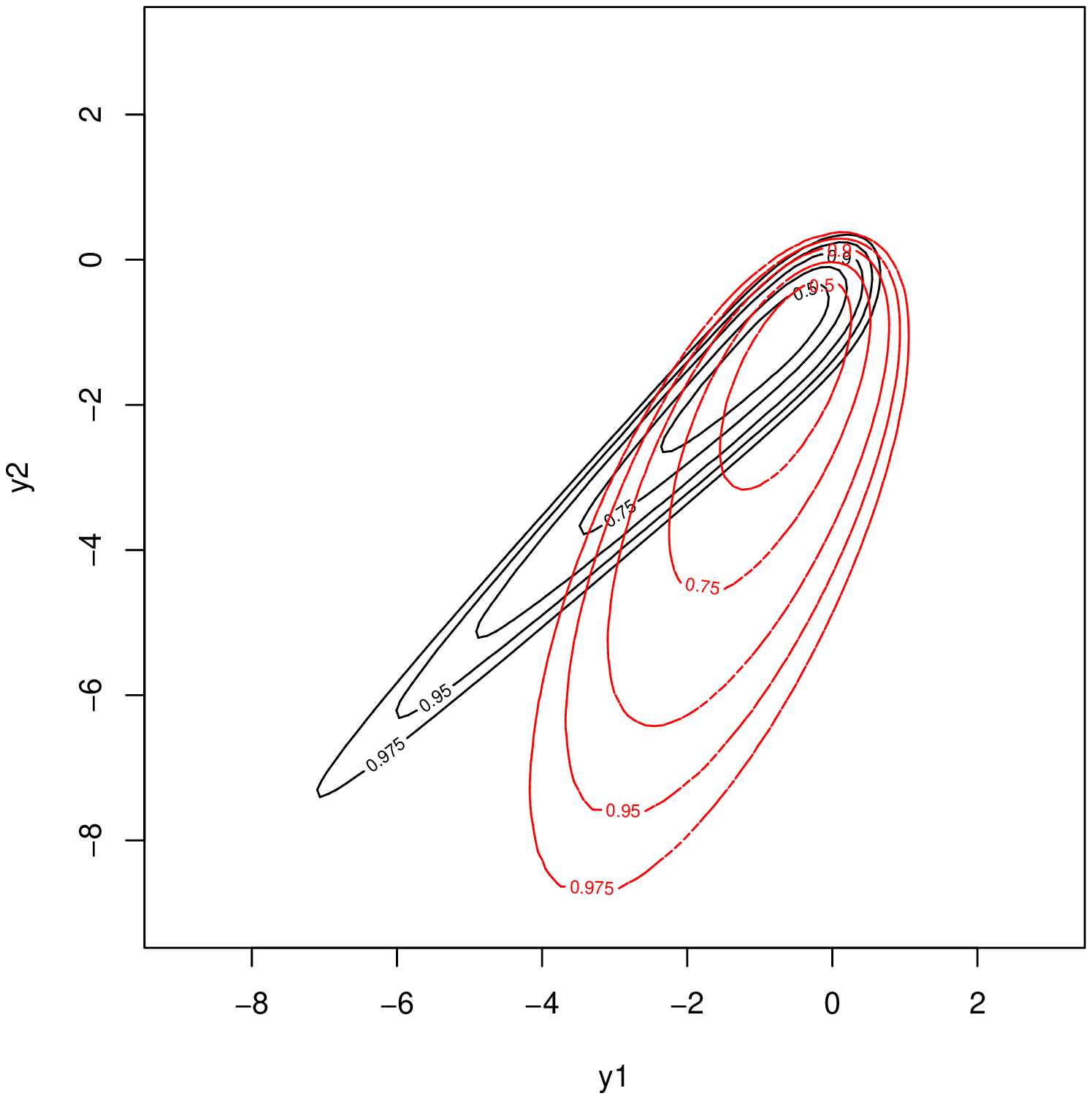} &
\hspace{-10mm} \includegraphics[scale=0.25]{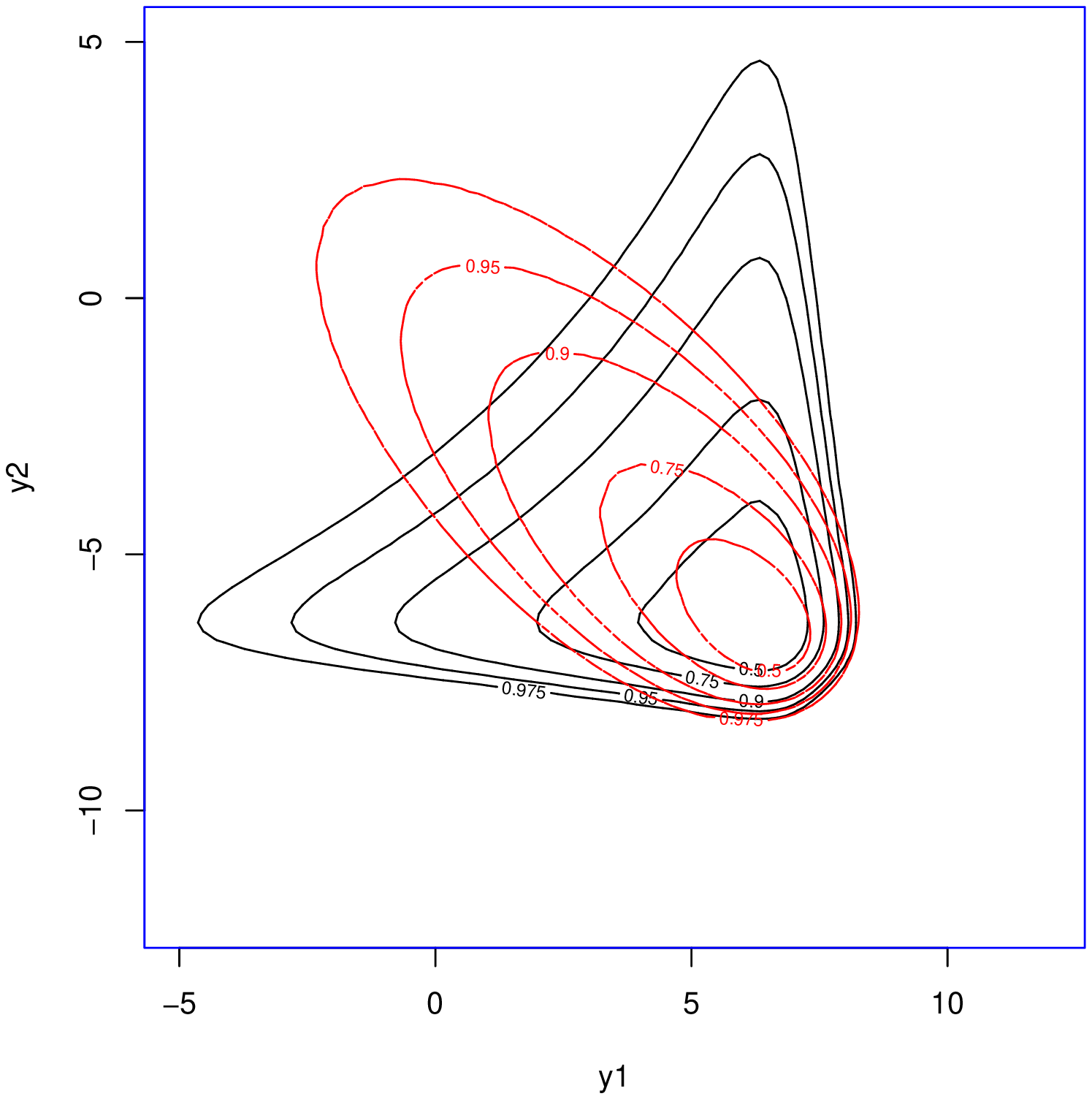} &
\hspace{-10mm} \includegraphics[scale=0.25]{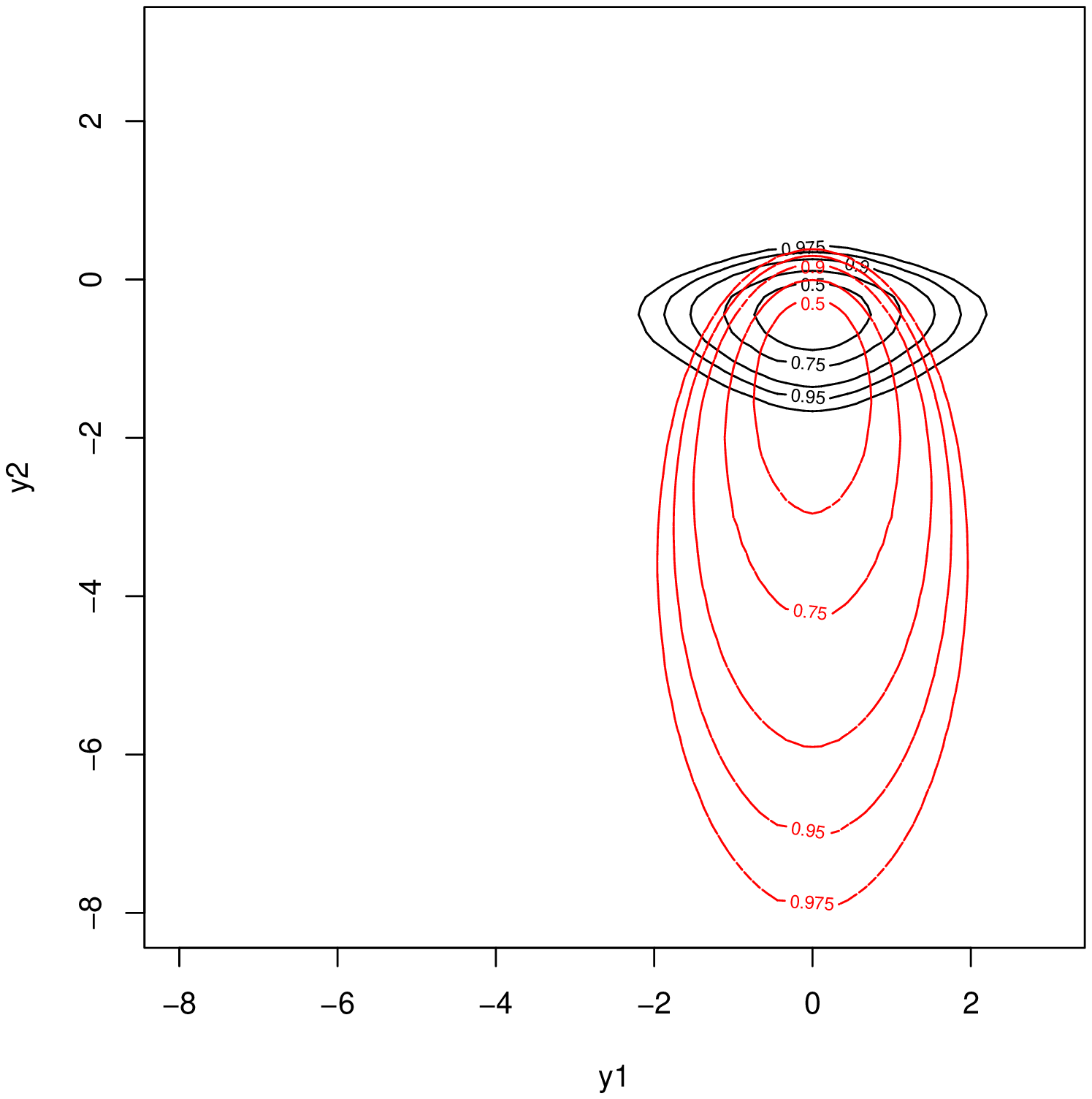} \\
(d) & (e) & (f)  \\
\hspace{-5mm} \includegraphics[scale=0.25]{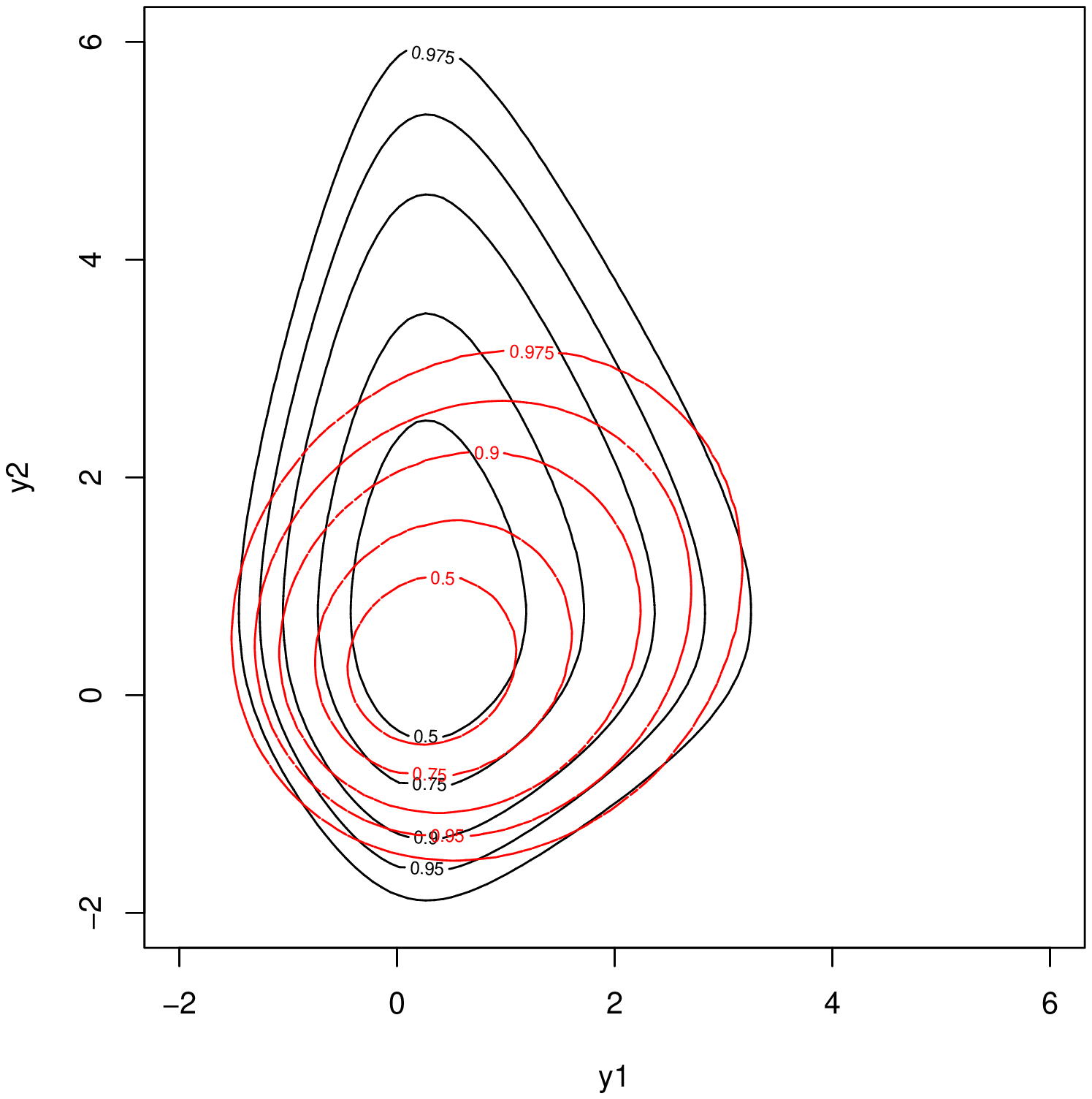} &
\hspace{-10mm} \includegraphics[scale=0.25]{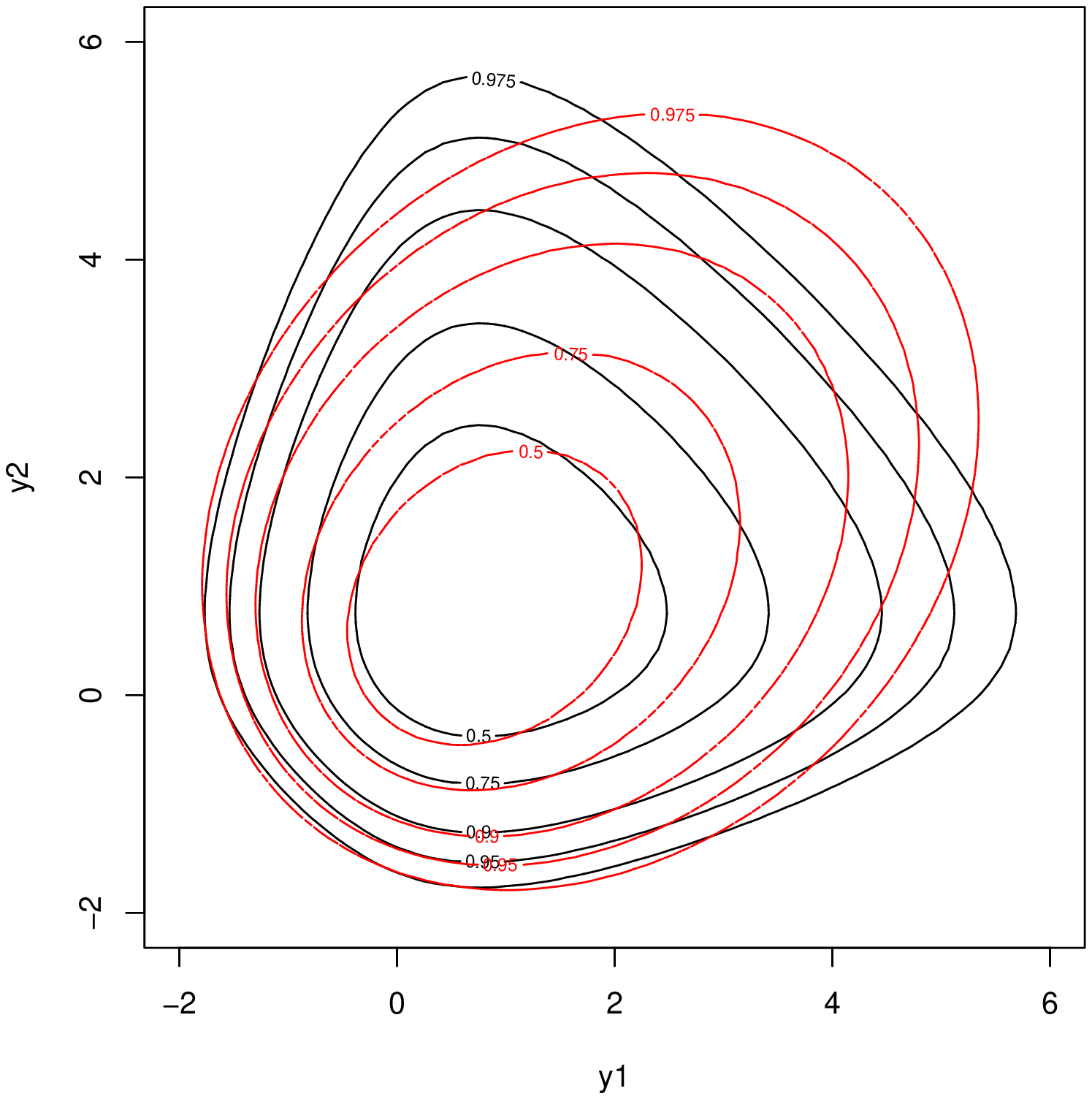} &
\hspace{-10mm} \includegraphics[scale=0.25]{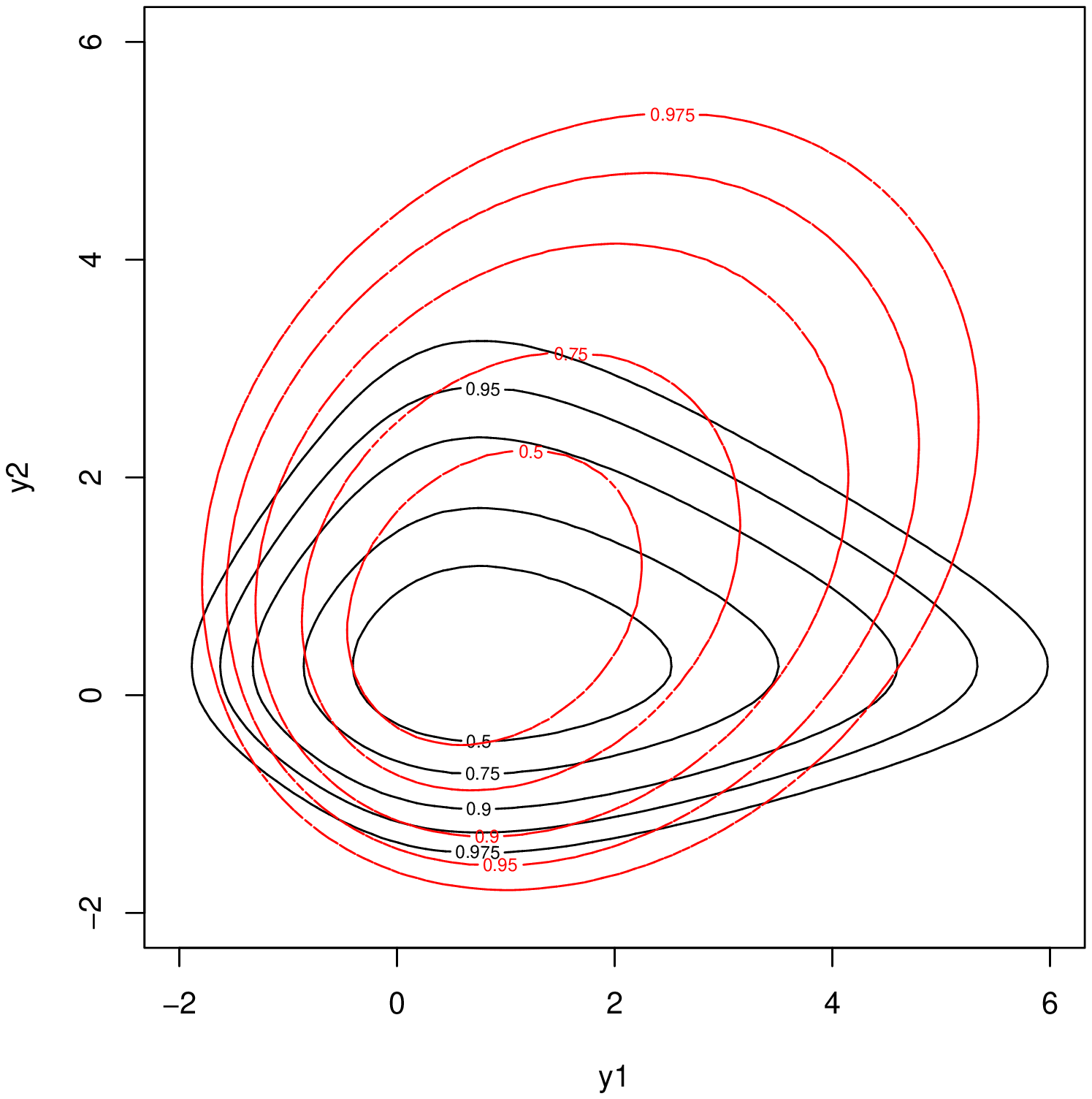} \\
(g) & (h) & (i)
\end{tabular}
\caption{\small Top and middle panels: contour plots of bivariate
multiple scaled NIG ($\lambda_1=\lambda_2=-1/2$) distributions
(solid lines) with $\mub=[0,0]^T$ and $\deltab=\{1,1\}$. The
difference with the standard multivariate NIG (red dashed lines)
is illustrated with univariate $\delta$ and $\gamma$ values taken
as the first respective component of the bivariate $\deltab$ and
$\gammab$. The (a-d) panels correspond to the same $\Sigmab$ built
from $\Ab=diag(3/2, 2/3)$ and $\xi=\pi/4$ with (a)
$\betab=[2,2]^T, \gammab=[1,1]^T$, (b) $\betab=[0,5]^T,
\gammab=[2,2]^T$, (c)$\betab=[0,-5]^T, \gammab=[2,2]^T$ and (d)
$\betab=[0,-5]^T, \gammab=[2,10]^T$ . The (e,f) panels correspond
to $\Sigmab=\Ib_2$  with (e) $\betab=[-2,2]^T, \gammab=[1,1]^T$,
(f) $\betab=[0,-5]^T, \gammab=[1,1]^T$. Bottom panels: contour
plots of various multiple scaled (solid lines) and standard (red
dashed lines) GH distributions all with $\Sigmab=\Ib_2$,
$\betab=[1,1]^T$, $\gammab=[2,2]^T$, $\deltab=[1,1]^T$ and (g)
$\lambdab=[-1/2,2]^T$, (h) $\lambdab=[-2,2]^T$, (i)
$\lambdab=[2,-1/2]^T$. } \label{fig:multnig}
\end{figure}

 The MSGH distribution (as defined in \eqref{eqn:mgh}) provides for very flexible
distributional forms. For illustration, in the bivariate case,
several contour plots of the multiple scaled NIG  ({\it i.e.} for all
$m$, $\lambda_m=-1/2$) are shown in Figure~\ref{fig:multnig} and
compared with the standard multivariate NIG. In this
two-dimensional setting, we use for $\Db$ a parameterisation via
an angle $\xi$ so that $D_{11}=D_{22}=\cos \xi$ and
$D_{21}=-D_{12}=\sin \xi$, where $D_{md}$ denotes the $(m,d)$
entry of matrix $\Db$.
 Similar to the
standard NIG the parameter $\betab$ measures asymmetry and its
sign determines the type of skewness. For the standard NIG the
contours are not necessarily elliptical and this is also the case
with the multiple scaled NIG. In the case of the multiple scaled
NIG additional flexibility is provided by allowing the parameter
$\gammab$ to be a vector of dimension $M$ instead of a scalar.
Keeping all $\delta_m$'s equal to the same $\delta$, this
vectorisation of $\gammab$ effectively allows each dimension to be
governed by different tail behaviour depending on the values of
$\gammab$ (see below).

Other multiple scaled and standard GH distributions are then also
illustrated in Figure~\ref{fig:multnig}. As shown in
Figure~\ref{fig:multnig}(g) and (i), changing $\lambda_m$ values
does not change much the shape of the contours but larger values
of $\lambda_m$ tend to produce heavier tails.

\subsubsection{Mean and covariance matrix}

Using the moments of the GIG distribution 
\citep[see][]{Jorgensen1982}, {\it i.e.}, if $W$ follows a ${\cal
GIG}(\lambda, \gamma,\delta)$ distribution, for all $r \in \Nset$,
\begin{equation}
E[W^r] =
\left(\frac{\delta}{\gamma}\right)^r\frac{K_{\lambda+r}(\delta\gamma)}{K_{\lambda}(\delta\gamma)},
\label{meanGIG}
\end{equation}
it follows from representation~\eqref{eqn:xwmulti} that when $\Yv$
follows a multiple scaled GH distribution,
\begin{eqnarray}
E[\Yv] & = & E[E[\Yv | \Wv]] =  \mub + \Db E[\Delta_{\wv}] \Ab \Db^{T} \betab \nonumber  \\
& = & \mub + \Db \hspace{1mm}
\text{diag}\bigg(\frac{\delta_m}{\gamma_m}\frac{K_{\lambda_m+1}(\delta_m\gamma_m)}{K_{\lambda_m}(\delta_m\gamma_m)}\bigg)
\Ab \Db^{T} \betab,\label{eqn:meanmgh}
\end{eqnarray}
where for short, we denoted by $\text{diag}(u_m)$ the
$M$-dimensional diagonal matrix whose diagonal components are
$\{u_1, \ldots, u_M\}$.

For the covariance matrix, we get,
\begin{align}
\textup{Var}[\Yv]  = & E[\textup{Var}[\Yv | \Wv]] + \textup{Var}[E[\Yv | \Wv]] \label{eqn:varmgh}  \\
 = & \Db E[\Delta_{\Wv}] \Ab \Db^{T} + \Db \Ab \hspace{1mm} \textup{Var} [\Delta_{\Wv} \Db^{T} \betab ] \Ab \Db^{T} \nonumber  \\
 =& \Db \hspace{1mm}
\text{diag}\bigg(\frac{\delta_mA_m}{\gamma_m}\frac{K_{\lambda_m+1}(\delta_m\gamma_m)}{K_{\lambda_m}(\delta_m\gamma_m)}
\bigg(1 + \frac{\delta_m}{\gamma_m}[\Db^{T}\betab]_m^2 \Ab_m \\ \notag
& \times \bigg(\frac{K_{\lambda_m+2}(\delta_m\gamma_m)}{K_{\lambda_m+1}(\delta_m\gamma_m)}
-
\frac{K_{\lambda_m+1}(\delta_m\gamma_m)}{K_{\lambda_m}(\delta_m\gamma_m)}\bigg)\bigg)\bigg)
\Db^{T} \nonumber
\end{align}
 For details of the mean and variance for the multiple
scaled NIG distribution see Appendix A.

As can be seen from~\eqref{eqn:varmgh}, the variance of the
multiple scaled GH takes a slightly complicated form with some
dependency on the skewness parameter $\betab$. This dependency is
also present in the variance, recalled below, of the standard
multivariate GH as given in (\ref{eqn:gh}),

\begin{equation}
\textup{Var}[\Yv_{GH}]=\frac{\delta}{\gamma}\frac{K_{\lambda+1}(\delta\gamma)}{K_{\lambda}(\delta\gamma)}\Sigmab
+
\frac{\delta^2}{\gamma^2}\bigg(\frac{K_{\lambda+2}(\delta\gamma)}{K_{\lambda}(\delta\gamma)}-\frac{K^2_{\lambda+1}(\delta\gamma)}{K^2_{\lambda}(\delta\gamma)}\bigg)
\Sigmab\betab^T\betab\Sigmab \; . \nonumber
\end{equation}

As noted recently by \cite{arellanovalleetal07} and \cite{leemclachlan12} an alternative form to allow separation between the skewness and the variance is provided in the case of the skew $t$ distribution as parameterised by \cite{sahuetal03}. Interestingly, other parameterisations of the skew-$t$ distribution
\citep{azzalinietal96} do not share this separation property.

A notable difference between the covariance structure of the
multiple scaled GH and the standard GH is that in the case of a
diagonal scale matrix $\Sigmab$, variates of the multiple scaled
GH are independent of each other. Interestingly, this is not the
case for the standard multivariate GH where the same latent factor
$W$ is shared across dimensions, and this effectively acts to
induce some degree of dependency between dimensions (although they
may be uncorrelated). A similar situation arises in the case of
other distributions with shared latent factors, for example the
standard  $t$-distribution. As mentioned previously, in the
multiple scaled GH case the latent factor $W$ is allowed to vary
independently across dimensions.

The tail behaviour of the multiple scaled GH is similar to the GH
with tails governed by a combined algebraic and exponential form
equivalent to $\delta
\text{exp}(\delta\gamma_m+[\Db^{T}\yv]_{(m)}[\Db^{T}\betab]_{(m)}
- \alpha_m q_m(\yv)) q_m(\yv)^{-1}$, where $q_m $ and $\alpha_m$
are defined in Equation~\eqref{eqn:mgh}. Hence, the multiple
scaled GH, like the GH distribution, is said to be {\it semi-heavy} tailed, which means that its tail behaviour is characterized by exponential instead of power decay.  Alternative parameterisations of
the GH permit the possibility of heavier tails \citep{aashaff06}.
The parameters $\gammab$, $\deltab$ and $\betab$ govern the tail
behaviour of the density with smaller values of $\gammab$ and
$\deltab$ implying heavier tails, and larger values lighter tails.
For our multiple scaled {\cal GH} distributions, when all $\delta_m, \gamma_m$
tend to infinity with $\delta_m/\gamma_m$ tending to 1, the
distribution tends to the multivariate Gaussian ${\cal
N}(\mub+\Sigmab \betab, \Sigmab)$. This is easily seen from the
characteristic function (see Section \ref{sec:cf}).

A difference between the tail behaviour of the GH and the multiple
scaled GH can also be seen in measures of the tail dependency
\citep{colesetal99}.  In applications, strong tail dependence is
important for modelling the dependency/association of potentially
extreme events ({\it e.g.} in finance, meteorology).  In
Figure~\ref{eqn:gig} we compare the tail dependency of the
Gaussian, $t$-distribution, standard GH and multiple scaled GH
using a $\chi(q)$ plot \citep{colesetal99} and simulated values
from each distribution with $\mub=[0,0]^T, \Sigmab=
\begin{pmatrix}
1 & 0.5 \\
0.5&1
\end{pmatrix}$
(equivalently $\Ab=diag(3/2, 1/2)$ and $\xi=\pi/4$) ,
$\betab=[0,0]^T, \gammab=\deltab=[1,1]^T$ (or $\nu=1$) and
$\lambdab=[-1/2,-1/2]^T \hspace{2mm} (\textup{NIG})$. The function
$\chi(q)$ can be interpreted as a quantile dependent measure of
dependence with $\chi(q)=0$ indicating  independence and $\chi(q)=1$ perfect dependence
(For further details see Appendix~D). Tail dependence is determined by the limit of $\chi(q)$ when $q$ tends to 1.


\begin{figure}[h!]
 \centering
  \includegraphics[scale=0.4]{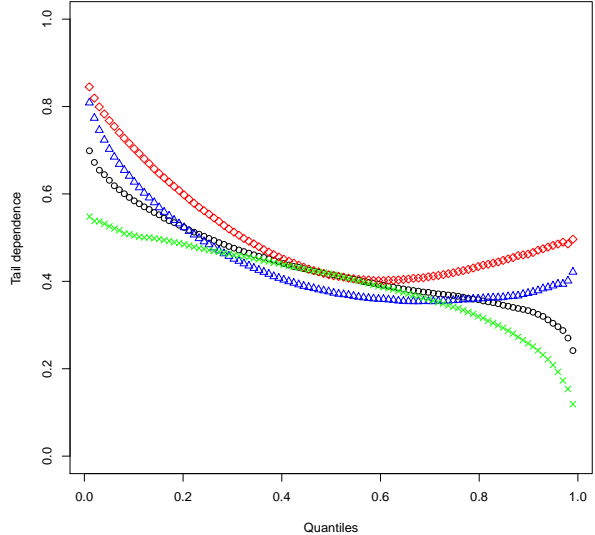}
  \caption{\footnotesize Comparison of tail dependence using $\chi(q)$. X-axis: quantiles levels.
   Y-axis: Estimate of $\chi(q)$, dashed lines indicate the $95 \%$ confidence interval. Gaussian distribution (Green), Standard NIG distribution (black), Multiple scaled
   NIG distribution (blue), $t$ distribution (Red) }
  \label{fig:taildependence}
\end{figure}

From Figure~\ref{fig:taildependence}, we see that the multiple
scaled NIG has stronger tail dependence than the standard NIG. By
comparison (and for reference), it is well known that the Gaussian
distribution has no tail dependence, and the
$t$-distribution has a stronger tail dependence than both the
Gaussian and the standard NIG.

\subsubsection{Characteristic function}

\label{sec:cf}

 Denote by $\phi_{\Yv}$ the characteristic function
of a random vector $\Yv$.  It follows from (\ref{eqn:multnigsim})
that, $\forall \tv \in \Rset^M, \; \phi_\Yv(\tv) = E[\exp(i \tv^T
\Yv)] = E[E[\exp(i \tv^T \Yv)|W]]  = \exp(i \tv^T \mub) \;
\prod\limits_{m=1}^M \phi_{W_m}(u_m(\tv))\; .$

where $u_m(\tv) =
[\Ab^{1/2}\Db^{T}\tv]_{m}([\Ab^{1/2}\Db^{T}\betab]_{m} +
\frac{i}{2}[\Ab^{1/2}\Db^{T}\tv]_{m})$ and $\phi_{W_m}$ is the
characteristic function of $W_m$.

In the Generalised Hyperbolic case $\phi_{W_m}$ is the
characteristic function of a 1-dimensional  ${\cal GIG}(\lambda_m,
\gamma_m, \delta_m)$ distribution, which is
\begin{equation}
\phi_{W_m}(t) = \left(\frac{\gamma_m}{\gamma_m -
2it}\right)^{\lambda_m}
\frac{K_{\lambda_m}(\delta_m\sqrt{\gamma_m^2 -
2it})}{K_{\lambda_m}(\delta_m\gamma_m)} \; .
\end{equation}

The particular case of the multiple scaled NIG follows easily by
setting $\lambda_m = -1/2$, which permits a  simpler form
\begin{equation}
\phi_{W_m}(t) = \text{exp}(\delta_m\gamma_m
-\delta_m\sqrt{\gamma_m^2 - 2it}) \; .
\end{equation}
The characteristic function is useful in practice for the computation of marginals as detailed in the next section.


\subsubsection{Marginals}
\label{margi}

Using~\eqref{eqn:xwmulti}, marginals are easy to sample from but
computing their pdfs involves, in general, numerical integration. An
efficient and simple algorithm to compute such marginal pdfs in
most cases can be derived according to \cite{Shephard1991}. The
derivation in \cite{Shephard1991} is based on the inversion
formula of the characteristic function which in the univariate
case is:
\begin{align}
f_{Y}(y) = & \frac{1}{2\pi} \int\limits_0^\infty (\exp(ity)
\phi_{Y}(-t)+\exp(-ity) \phi_{Y}(t)) dt \\ \notag
= & \frac{1}{\pi}
\int\limits_0^\infty Re(\exp(-ity) \phi_{Y}(t)) dt \;
\end{align}
using the hermitian property of characteristic functions
$\phi_{Y}(-t)= {\overline{\phi_{Y}(t)}}$ (the over line means the
complex conjugate).

\begin{figure}[h!]
 \centering
 \begin{tabular}{c}
  \includegraphics[scale=0.3]{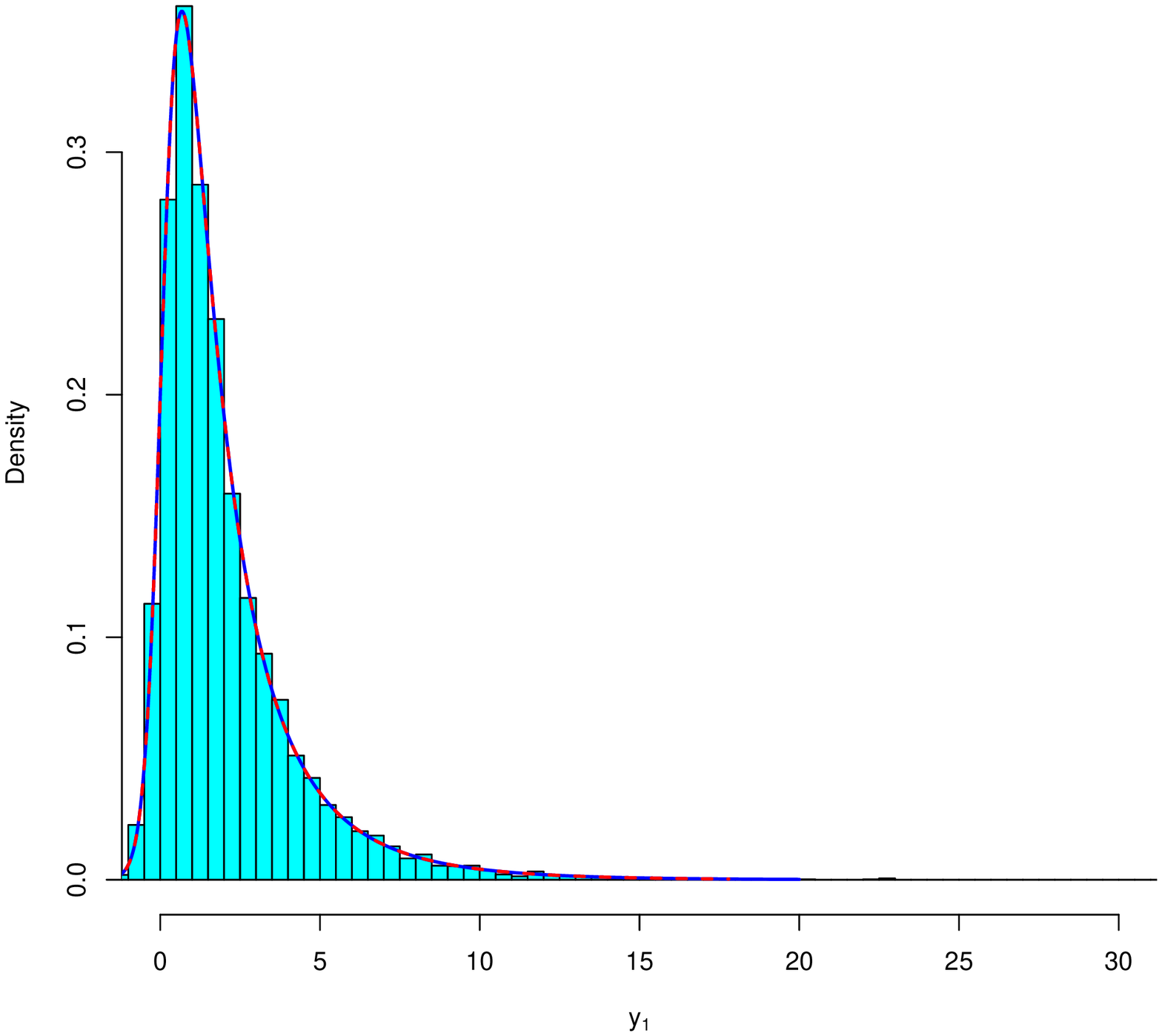}
  \includegraphics[scale=0.3]{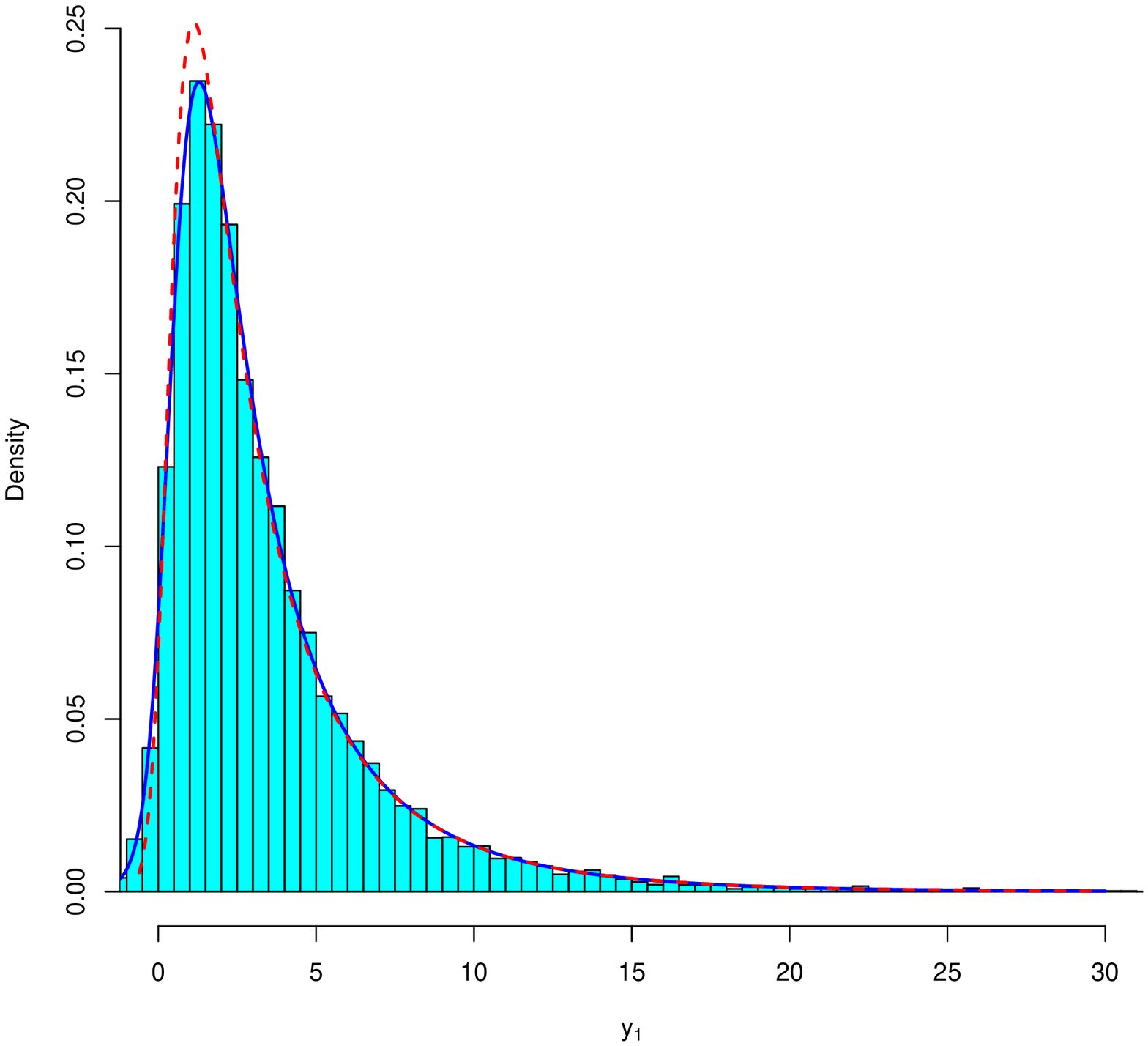}
    \end{tabular}
  \caption{\footnotesize Histogram and density plots of the marginal $\Yv_1$ of a bivariate NIG distribution
   with $\mub=[0,0]^T$, $\gammab=\deltab=\betab=[2,2]^T$, and (left) diagonal $\Sigmab$ with diagonal entries equal to 1 or
    (right) $\Sigmab$ with diagonal entries equal to 1 and other entries to 0.5. Histograms and blue solid lines denote the multiple scaled NIG
         and red dashed lines the standard NIG. \label{fig:unimarg}}
\end{figure}

As an illustration, Figure~\ref{fig:unimarg} shows plots of the
pdf of some 1-D marginals and a comparison with 1-D NIG
distributions.
From Figure~\ref{fig:unimarg}  we can see that the marginals of
the proposed multiple scaled NIG (MSNIG) distribution deviate
slightly from the standard NIG distribution according to the
specification of $\Sigmab$. The marginals of the  MSNIG
distribution are exactly 1-D standard NIG distributions in the
diagonal scale matrix case.


\vspace{.3cm}
For marginals of dimension greater than 1, we can also easily derive the characteristic function and use a simple multidimensional inversion formula.
Let ${\cal I}$ be  a subset of $\{1, \ldots, M\}$ of size $I$ and write $\Yb_{\cal I}=\{Y_m, m \in {\cal I}\}$ and $\tb_{\cal I}=\{t_m, m \in {\cal I}\}$.
 The characteristic function of the marginal variable $\Yb_{\cal I}$ is
\begin{equation}
\phi_{\Yb_{\cal I}}(\tb_{\cal I}) = \prod\limits_{m \in {\cal I}}
\exp(i t_m \mu_m) \; \prod\limits_{d=1}^M
\phi_{W_d}(u_{d}(\tb_{\cal I})) \; ,
\end{equation}
with $u_{d}(\tb_{\cal I}) = (\sum\limits_{m \in {\cal I}} t_m
[\Db\Ab^{1/2}]_{md}\; [\Ab^{1/2}\Db^{T}\betab]_{d}) + \frac{i}{2}
(\sum\limits_{m\in {\cal I}} t_m [\Db\Ab^{1/2}]_{md})^2\; .$

 It follows that the density of $\Yb_{\cal I}$ via the
multidimensional inversion formula (see {\it e.g.}
\cite{Shephard1991}) is:
\begin{equation}
f_{\Yb_{\cal I}}(\yb_{\cal I}) = (2\pi)^{-I} \int\limits_{-\infty}^\infty \ldots
\int\limits_{-\infty}^\infty \exp(-i \tb_{\cal I}^T \yb_{\cal I})
\; \phi_{\Yb_{\cal I}}(\tb_{\cal I}) \; d\tb_{\cal I}
\end{equation}

When $I=2$, and decomposing $\Rset^2$ into four quadrants,
\begin{equation}
f_{\Yb_{\cal I}}(\yb_{\cal I}) = 2 \; (2\pi)^{-2}
\int\limits_{0}^\infty \; \int\limits_{-\infty}^\infty Re(\exp(-i
\tb_{\cal I}^T \yb_{\cal I}) \; \phi_{\Yb_{\cal I}}(\tb_{\cal I}))
\; d\tb_{\cal I} .
\label{eqn:margbiv}
\end{equation}
This formula also generalizes easily in higher dimensions.

\begin{figure}[h!]
 \centering
  \includegraphics[scale=0.4]{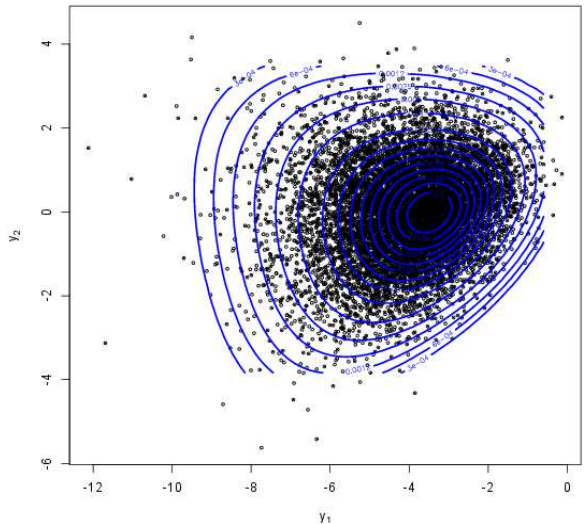}
  \caption{\footnotesize $[Y_1,Y_2]^T$ distribution when $[Y_1,Y_2,Y_3]^T$ follows a multiple scaled trivariate NIG distribution with $\mub=[0,0,0]^T$,
  $\gammab=\deltab=[3,3,3]^T$, $\betab=[-6,2,2]^T$
   and
  $\Sigmab$ so that its diagonal entries
are 1 and other entries are 0.5. Contours are superimposed on
points sampled from the distribution using equation
\eqref{eqn:xwmulti}.\label{fig:bivmarg}}
\end{figure}

For illustration, Figure \ref{fig:bivmarg} shows the bivariate
marginal $[Y_1, Y_2]^T$ for a 3 dimensional $[Y_1, Y_2, Y_3]^T$
following a  MSNIG distribution with $\mub=[0,0,0]^T$,
$\gammab=\deltab=[3,3,3]^T$, $\betab=[-6,2,2]^T$ and $\Sigmab$ so
that its diagonal entries are 1 and other entries are 0.5. It is
clear from the shape of the contours that this bivariate marginal
takes a slightly different shape to a bivariate standard NIG
distribution.

\section{Maximum likelihood estimation of parameters}\label{sec:ML}

In this section, for illustration we outline an EM approach to
estimate the parameters  of the multiple scaled NIG distribution
as it appears to be the most popular case of the GH family used in
applications especially in finance.  As noted also by
\citep{protassov04,barndorffnielsen97}, for the GH distribution it
can be very difficult to show a significant difference between
different values of $\lambda$ due to the flatness of the
likelihood and computational difficulties arise in some cases
where the likelihood can be infinite.  For these reasons we
outline the particular case of allowing all $\lambda_m$'s to be
fixed but we note that it is relatively straightforward to extend
our proposed approach to the more general case. Also for
identifiability reasons, we set all $\delta_m$'s to the same
$\delta$ value so that the parameters to estimate in the multiple
scaled NIG case are $\Psib=\{\mub, \Db, \Ab, \betab, \gammab,
\delta\}$ with $|\Ab|=1$.

Estimation of most of the parameters for the multiple scaled NIG
distribution is relatively straightforward but the separate
estimation of $\Db$ and $\Ab$ requires an additional minimization
algorithm based on the Flury and Gautschi algorithm
\citep{Flury1984,FluryGautschi1986}. Similar difficulties are also
encountered in Gaussian model-based clustering
\citep{CeleuxGovaert1995} for some of the proposed models.

Let us consider  an {\it i.i.d} sample $\yv=\{\yv_1,
\ldots,\yv_N\}$ of the multiple scaled NIG distribution defined in
\eqref{eqn:mnig}. As in the standard NIG distribution case \citep{Karlis2002}, a
convenient computational advantage of the EM approach is to view
the weights as an additional missing variable $\Wb$.  The observed
data $\yv$ are seen as being incomplete and additional missing
weight variables $\Wb_{1} \ldots \Wb_{N}$ with for $i \in \{1
\ldots N\}$, $\Wb_{i} = [W_{i1} \ldots W_{iM}]^T$ are introduced.
These weights are defined so that $\forall i \in \{1\ldots N\}$:
\begin{eqnarray}
\Yv_i | \Wb_{i}= \wb_i &\sim& \mathcal{N}_M(\mub+  \Db
\Deltab_{\wv_i} \Ab \Db^{T} \betab, \Db \Deltab_{\wv_i} \Ab
\Db^{T}) \label{repre}\\
\mbox{and} \quad \Wb_{i} &\sim& {\cal IG}(\gamma_1,\delta) \otimes
\dots \otimes {\cal IG}(\gamma_M,\delta) \nonumber
\end{eqnarray}
 where $\Deltab_{\wv_i} = \text{diag}(w_{i1}, \ldots, w_{iM})$ .

As a way of circumventing the restriction that the determinant
$|\Ab|=1$ in the M-step, representation (\ref{repre}) above  can
be rewritten equivalently as,
\begin{eqnarray}
\Yv_i|\Wb_i=\wb_i & \sim & \mathcal{N}_M(\mub+\Db\Deltab_{\wv_i}\Db^{T}\tilde{\betab},\Db\Deltab_{\wv_i}\tilde{\Ab}\Db^{T}) \label{reprebis} \\
\Wb_i & \sim & {\cal IG}(\tilde{\gamma}_{1},1) \otimes \dots
\otimes {\cal IG}(\tilde{\gamma}_{M},1) \nonumber
\end{eqnarray}
where
\begin{eqnarray*}
\tilde{\Ab} &=& \delta^2 \; \Ab \\
\tilde{\betab} &=& \Db\tilde{\Ab}\Db^{T}\betab \\
\tilde{\gammab} &=& \delta \; \gammab
\end{eqnarray*}
and $\tilde{\Ab} $ is now a general (positive definite) diagonal
matrix. Note that in the location term in the definition 
above (\ref{reprebis}), $\Db\Deltab_{\wv_i}\Db^{T}\tilde{\betab}=\Db\Deltab_{\wv_i}
\tilde{\Ab} \Db^{T} \betab$.

\subsection{E step}

\label{estep}

 At iteration $(r)$ with $\psib^{(r)}$ being the
current parameter value, the E-step leads to the computation for
all $i=1,\dots,N$, of the missing variables posterior
distributions $p(\wb_i|\yv_i;\psib^{(r)})$. It consists then of
calculating $p(\wb_i|\yv_i;\psib^{(r)}) \propto
p(\yv_i|\wb_i;\psib^{(r)})p(\wb_i;\psib^{(r)})$ which can be shown
\citep[see Appendix of][]{KarlisSantourian2009} to follow a
Generalised Inverse Gaussian  distribution (see
definition~\eqref{eqn:gig}). In our case, and assuming the
$\Wb_i$'s are independent we have,
\begin{equation}
p(\wb_i|\yv_i;\psib^{(r)}) = \prod_{m=1}^{M} {\cal GIG}(w_{im};
-1,\hat{\alpha}_{m}^{(r)}, \phi_{im}^{(r)}),
\end{equation}
where

\begin{eqnarray*}
\phi_{im}^{(r)} & = & \displaystyle \sqrt{1+\frac{[\Db^{(r)T}(\yv_i-\mub^{(r)})]^{2}_{(m)}}{\tilde{\Ab}_m^{(r)}}} \\
\hat{\alpha}_{m}^{(r)} & = & \displaystyle
\sqrt{\tilde{\gamma}_{m}^{(r)2}+\frac{[\Db^{(r)T}\tilde{\betab}^{(r)}]^{2}_{m}}{\tilde{\Ab}_{m}^{(r)}}}
\; .
\end{eqnarray*}

As all moments of a Generalised Inverse Gaussian distribution
exist (see (\ref{meanGIG})), it follows that we have closed form expressions
for the following quantities needed in the E-step,

\begin{eqnarray*}
s_{im}^{(r)} & = & E[W_{im}|\yv_{i};\psib^{(r)}] = \frac{\phi_{im}^{(r)} K_{0}(\phi_{im}^{(r)}\hat{\alpha}_{m}^{(r)})}{\hat{\alpha}_{m}^{(r)} K_{-1}(\phi_{im}^{(r)}\hat{\alpha}_{m}^{(r)})} \\
t_{im}^{(r)} & = & E[W^{-1}_{im}|\yv_{i};\psib^{(r)}] =
\frac{\hat{\alpha}_{m}^{(r)}
K_{-2}(\phi_{im}^{(r)}\hat{\alpha}_{m}^{(r)})}{\phi_{im}^{(r)}K_{-1}(\phi_{im}^{(r)}\hat{\alpha}_{m}^{(r)})}\;.
\end{eqnarray*}

Note that equivalently $K_{-1}=K_{1}$ and $K_{-2}= K_2$. The
Bessel function can be numerically evaluated in most statistical
packages.  All computations in this paper were undertaken using R
\citep{Rsoftware}.

\subsection{M step}

For the updating of $\psib$, the M-step consists of two independent steps for ($\mub,\Db,\tilde{\Ab},\tilde{\betab}$) and $\tilde{\gammab}$,

\begin{align}
(\mub,\Db,\tilde{\Ab}, \tilde{\betab})^{(r+1)}  = &
\arg\max_{\mub,\Db,\Ab,\betab} \sum_{i=1}^{N} E[\text{log}
\hspace{0.5mm} p(\yv_i,|\Wv_i; \mub, \Db,
\tilde{\Ab},\tilde{\betab})|\yv_i,\psib^{(r)}] \label{eqn:mstep1}
\\ \nonumber  = & \arg\max_{\mub,\Db,\Ab,\betab} \bigg\{
\sum_{i=1}^{N} - \frac{1}{2}\text{log}|\tilde{\Ab}|
-\frac{1}{2}(\yv_i-\mub-\Db\Sb_i^{(r)}\Db^{T}\tilde{\betab})^{T} \\ \notag
\times & \hspace{2mm} \Db\tilde{\Ab}^{-1}\Tb_i^{(r)}\Db^{T}(\yv_i-\mub-\Db\Sb_i^{(r)}\Db^{T}\tilde{\betab})\bigg\}
\end{align}

and
\begin{eqnarray}
\tilde{\gammab}^{(r+1)} & = & \arg\max_{\tilde{\gammab}}
\sum_{i=1}^{N}\sum_{m=1}^{M} E[\text{log} \hspace{0.5mm}
p(W_{im};\tilde{\gamma}_{m},1)|\yv_i,\psib^{(r)}]
\label{eqn:mstep2} \\ \nonumber & = & \arg\max_{\tilde{\gammab}}
\bigg\{ \sum_{i=1}^{N}\sum_{m=1}^{M} \tilde{\gammab}_{m} -
\frac{1}{2}\tilde{\gammab}_{m}^{2}s_{im}^{(r)})\bigg\}
\end{eqnarray}
where $\Tb_i^{(r)}=\text{diag}(t_{i1}^{(r)},\dots,t_{iM}^{(r)})$
and $\Sb_i^{(r)}=\text{diag}(s_{i1}^{(r)},\dots,s_{iM}^{(r)})$ and
ignoring constants.

The optimization of these steps leads to the following update equations.

\noindent {\bf Updating $\mub$.} It follows from \eqref{eqn:mstep1} that for fixed $\Db$ and $\Ab$ (ignoring constants)

\begin{equation}
\mub^{(r+1)} = \arg\min_{\mub} \bigg\{ \sum_{i=1}^{N}
(\yv_i-\mub-\Db\Sb_i^{(r)}\Db^{T}\tilde{\betab})^{T}
\Db\tilde{\Ab}^{-1}\Tb_i^{(r)}\Db^{T}(\yv_i-\mub-\Db\Sb_i^{(r)}\Db^{T}\tilde{\betab})\bigg\}
\end{equation}

which by fixing $\Db$ to the current estimation $\Db^{(r)}$, leads to

\begin{align*}
\mub^{(r+1)} =&
\bigg(\frac{\sum_{i=1}^{N}\Tb_{i}^{(r)}\Db^{(r)T}}{N}-N \;
(\sum_{i=1}^{N}\Sb_{i}^{(r)})^{-1}\bigg)^{-1} \\ \notag 
& \times \bigg(\frac{\sum_{i=1}^{N}
\Tb_{i}^{(r)}\Db^{(r)T}\yv_i}{N}-\sum_{i=1}^{N}\yv_i \;
(\sum_{i=1}^{N}\Sb_{i}^{(r)})^{-1}\bigg)
\end{align*}

\noindent {\bf Updating $\tilde{\betab}$.}  To update $\tilde{\betab}$ we have to minimize the following quantity,

\begin{equation}
\tilde{\betab}^{(r+1)} = \arg\min_{\tilde{\betab}} \bigg\{
\sum_{i=1}^{N}
(\yv_i-\mub-\Db\Sb_i^{(r)}\Db^{T}\tilde{\betab})^{T}
\Db\tilde{\Ab}^{-1}\Tb_i^{(r)}\Db^{T}(\yv_i-\mub-\Db\Sb_i^{(r)}\Db^{T}\tilde{\betab})\bigg\}
\end{equation}

which by fixing $\Db$ and $\mub$ to their current estimations $\Db^{(r)}$ and $\mub^{(r+1)}$, leads to

\begin{eqnarray*}
\tilde{\betab}^{(r+1)}=\Db^{(r)}(\sum_{i=1}^{N}\Sb_{i}^{(r)})^{-1}\Db^{(r)T}\sum_{i=1}^{N}(\yv_i-\mub^{(r+1)})
\end{eqnarray*}

\noindent {\bf Updating $\Db$.} Using the equality $x^T \Sb
x=\text{trace}(\Sb xx^T)$  for any matrix $\Sb$, it follows that
for fixed $\tilde{\Ab}$ and $\mub$,  $\Db$ is obtained by
minimizing 
\begin{align*}
\Db^{(r+1)} = & \arg\min_{\Db} \bigg\{ \sum\limits_{i=1}^N
\text{trace}(\Db \Tb_{i}^{(r)}
\tilde{\Ab}^{(r)-1}\Db^{T}\Vb_{i})+\sum\limits_{i=1}^N
\text{trace}(\Db \Sb_{i}^{(r)} \tilde{\Ab}^{(r)-1}\Db^{T}\Bb_{i})
\\ \notag 
& - 2\sum\limits_{i=1}^N\text{trace}(\Db\tilde{\Ab}^{(r)-1}\Db^{T}\Cb_{i})
\bigg\}
\end{align*}
 where $\Vb_{i}=(\yv_i- \mub^{(r+1)}) (\yv_i
-\mub^{(r+1)})^T, \Bb_{i}=\tilde{\betab}^{(r+1)}
\tilde{\betab}^{(r+1)T}$,
$\Cb_{i}=(\yv_i- \mub^{(r+1)})\tilde{\betab}^{(r+1)T}$.

Using current values $\mub^{(r+1)}, \betab^{(r+1)}$ and
$\tilde{\Ab}^{(r)}$, the parameter $\Db$ can be updated using an
algorithm derived from Flury and Gautschi (see
\citet{CeleuxGovaert1995}) which is outlined in Appendix B.

\vspace{5mm}

\noindent \textbf{Updating $\tilde{\Ab}$.} To update $\tilde{\Ab}$
we have to minimize the following quantity
\begin{align*}
\tilde{\Ab}^{(r+1)} = & \arg\min_{\tilde{\Ab}} \bigg\{ \sum\limits_{i=1}^N \text{trace}(\Db^{(r+1)} \Tb_{i}^{(r)} \tilde{\Ab}^{-1}\Db^{(r+1)T}\Vb_{i})\\ \notag
& + \sum\limits_{i=1}^N \text{trace}(\Db^{(r+1)} \Sb_{i}^{(r)} \tilde{\Ab}^{-1}\Db^{(r+1)T}\Bb_{i}^{(r)}) - \\
 & 2\sum\limits_{i=1}^N\text{trace}(\Db^{(r+1)} \tilde{\Ab}^{-1}\Db^{(r+1)T}\Cb_{i})+ N \log|\tilde{\Ab}| \bigg\} \\
 = & \arg\min_{\tilde{\Ab}} \bigg\{
\text{trace}\left[\sum\limits_{i=1}^{N}(T_{i}^{(r)1/2}\Db^{(r+1)T}\Vb_{i}\Db^{(r+1)} \Tb_{i}^{(r)1/2} + \right. \\
& \left. \Sb_{i}^{(r)1/2}\Db^{(r+1)T}\Bb_{i}\Db^{(r+1)} \Sb_{i}^{(r)1/2}-
\Db^{(r+1)T}(\Cb_{i}+\Cb_{i}^{T})\Db^{(r+1)})\tilde{\Ab}^{-1} \right] \\ \notag
& + N \text{log}|\tilde{\Ab}| \bigg\} \\
=& \arg\min_{\tilde{\Ab}}
\bigg\{\text{trace}((\sum\limits_{i=1}^{N} \Mb_i) \;
\tilde{\Ab}^{-1})+N\text{log}\tilde{\Ab}\bigg\}
\end{align*}
where $\Mb_i=\Tb_{i}^{(r)1/2}\Db^{(r+1)T}\Vb_{i}\Db^{(r+1)}
\Tb_{i}^{(r)1/2} + \Sb_{i}^{(r)1/2}\Db^{(r+1)T}\Bb_{i} \Db^{(r+1)}
 \Sb_{i}^{(r)1/2}-\Db^{(r+1)T}(\Cb_{i}+\Cb_{i}^{T})\Db^{(r+1)}$ and $\Mb_i$ is a symmetric positive definite
 matrix.

We can use the following corollary (see Corollary A-2 in
\citet{CeleuxGovaert1995}) with $\Sb=\sum\limits_{i=1}^{N}\Mb_i$.

\textit{\textbf{Corollary 3.2}: \label{cor:updateA} The $M \times
M$ diagonal matrix $\Ab$ minimizing $\text{trace}(\Sb\Ab^{-1})$ +
$\alpha$\text{log}$|\Ab|$ where $\Sb$ is a $M\times M$ symmetric
definite positive matrix and $\alpha$ is a positive real number is
$\Ab = \frac{\text{diag}(\Sb)}{\alpha}$}

By setting $\Db$ and $\mub$ to their current estimations
$\Db^{(r+1)}$ and $\mub^{(r+1)}$ we then get,

\begin{equation}
\tilde{\Ab}^{(r+1)} =\frac{\text{diag}(\Sb)}{N}
\end{equation}
where
\begin{align}
\Sb = &
\sum\limits_{i=1}^{N}(\Tb_{i}^{(r)1/2}\Db^{(r+1)T}\Vb_{i}\Db^{(r+1)}
\Tb_{i}^{(r)1/2}+\Sb_{i}^{(r)1/2}\Db^{(r+1)T}\Bb_{i}\Db^{(r+1)}
\Sb_{i}^{(r)1/2} \\ \notag
& -\Db^{(r+1)T}(\Cb_{i}+\Cb_{i}^{T})\Db^{(r+1)}) \;
.
\end{align}

Equivalently, for all $m$

\begin{align}
\tilde{\Ab}^{(r+1)}_{m}= & \frac{1}{N}\sum\limits_{i=1}^{N}\bigg([\Db^{(r+1)T}(\yv_i-\mub^{(r+1)})]^{2}_m
t^{(r)}_{im}+[\Db^{(r+1)T}\tilde{\betab}^{(r+1)}]^{2}_m
s^{(r)}_{im} \\ \notag
& -2[\Db^{(r+1)T}(\yv_i-\mub^{(r+1)})]_m[\Db^{(r+1)T}\tilde{\betab}^{(r+1)}]_m\bigg)
\end{align}

 \noindent {\bf Updating $\tilde{\gammab}$.}  It
follows from \eqref{eqn:mstep2} that to update $\tilde{\gammab}$
we have to minimize,

\begin{equation}
\tilde{\gammab}_{m}^{(r+1)} = \arg\min_{\tilde{\gammab}} \bigg\{
\sum_{i=1}^{N}\sum_{m=1}^{M}
\frac{1}{2}\tilde{\gammab}_{m}^{2}s_{im}^{(r)} -
\tilde{\gammab}_{m})\bigg\} \label{eqn:mstep2bis}
\end{equation}

which for all $m = 1,\dots, M$ leads to,

\begin{eqnarray*}
\tilde{\gammab}_{m}^{(r+1)} &=& \displaystyle
\frac{N}{\sum_{i=1}^{N} s_{im}^{(r)}} \; .
\end{eqnarray*}

 \noindent{\bf Updating constrained $\tilde{\gammab}$.} Similar updating equations can be
 easily derived when $\tilde{\gammab}$ is assumed to be equal for several dimensions.
 If we assume that for all $m$, $\tilde{\gamma}_m=\tilde{\gamma}$ then

 \begin{eqnarray*}
 \tilde{\gammab}^{(r+1)} &=& \displaystyle \frac{NM}{\sum_{i=1}^{N}\sum_{m=1}^{M} s_{im}^{(r)}}
 \end{eqnarray*}

It is also quite easy to extend the above equation to the case
where $\tilde{\gammab}$ is assumed to be equal for only some of
the dimensions. For either case, model choice criteria could be
used to justify the appropriateness of the assumed parameter space
for $\tilde{\gammab}$.

Eventually, to transform the estimated parameters back to their
original form we can take  $
 \delta = |\tilde{\Ab}|^{\frac{1}{2M}},
 \gamma_{m}  = \tilde{\gammab}_{m}/\delta,
 \betab  =  \Db\tilde{\Ab}^{-1}\Db^{T}\tilde{\betab}$
and $\Ab  = \tilde{\Ab}/|\tilde{\Ab}|^{1/M}$.

\subsection{Mixture of multiple scaled NIG distributions}\label{sec:mix}

The previous results can be extended to cover the case of
$K$-component mixture of multiple scaled NIG distributions. With
the usual notation for the proportions $\pib=\{\pi_1, \ldots,
\pi_K\}$ and $\psib_k =\{\mub_k, \Db_k,\Ab_k, \betab_k,
\gammab_k,\delta_k \}$ for $k=1 \ldots K$, we consider, $$
p(\yv;\phib) = \sum_{k=1}^{K}\pi_k {\cal MSNIG}(\yv;\mub_k,\Db_k,\Ab_k, \betab_k, \gammab_{k},\delta_k)$$ where
$k$ indicates the $k$th component of the mixture and
$\phib=\{\pib, \psib\}$ with $\psib=\{\psib_1, \ldots \psib_K\}$
the mixture parameters. In the EM framework, an additional
variable $\Zb$ is introduced to identify the missing class labels,
where $\{Z_1,\dots,Z_N\}$ define the component of origin of the
data $\{\yv_1,\dots,\yv_N\}$. In the light of the characterization
of multiple scaled distributions, an equivalent modelling is:
 $\forall i \in \{1\ldots N\}, \\
 \quad \Yv_i | \Wb_{i}= \wb_i, Z_i= k \sim \mathcal{N}_M(\mub_k+  \Db_k \Deltab_{\wv_{i}}
  \Ab_k \Db_k^{T}\betab_k,\Db_k\Deltab_{\wv_{i}} \Ab_k \Db_k^{T})$
 and
$\Wb_{i} | Z_i=k \sim  {\cal IG}(\gamma_{1k},\delta_k) \otimes
\dots \otimes {\cal IG}(\gamma_{Mk},\delta_k)  \; ,$ where
$\Deltab_{\wv_i} = \text{diag}(w_{i1}, \ldots, w_{iM})$. Inference
using the EM algorithm with two sets of missing variables
$\Zb=\{Z_1, \ldots, Z_N\}$ and $\Wb=\{\Wb_1, \ldots, \Wb_N\}$ to
fit such mixtures, is similar to the individual ML estimation (see
Appendix C).

As the results of the EM algorithm can be particularly sensitive
to initial values \citep{karlisetal03}, for the results to follow
we used a number of approaches to generate different initial
values for parameters, including the use of random partitions,
$k$-means and trimmed $k$-means \citep{garciaetal99}. Often the
most successful strategy found was by estimating $\mub_k$, $\Db_k$
and $\Ab_k$ using the results from a trimmed $k$-means clustering
(with $\betab_k=0$) and setting $\gamma_{km}=\delta_k=1$ for all
$k=1\ldots K$ and $m=1\ldots M$. The computational speed of the EM
algorithm for the MSNIG distribution is comparable to the standard NIG case with the
exception that the update of $\Db$ can be slow for high
dimensional applications as the Flury and Gautschi algorithm
involves sequentially updating every pair of column vectors of
$\Db$.  A more global approach to the update of $\Db$ has been
proposed recently by \cite{browneetal13} which has the potential
to significantly speed up the computation time.

\section{Applications of multiple scaled NIG distributions}\label{sec:appl}

In this section we use simulated data and present applications of the multiple scaled NIG distribution on two real datasets to demonstrate its flexibility in analysing skewed multivariate data.

\subsection{Simulated data}

For this example, we simulated data from a mixture of MSNIG distributions and assessed the classification performance of the MSNIG compared to the standard NIG. The classification performance of the MSNIG compares favourably to the standard NIG with the MSNIG better able to capture the heavy tails of the two clusters. For details see Appendix A of the Supplementary Materials.

\subsection{Petroleum data}

\begin{figure}[htpb]
\begin{tabular}{cc}
\includegraphics[scale=0.3]{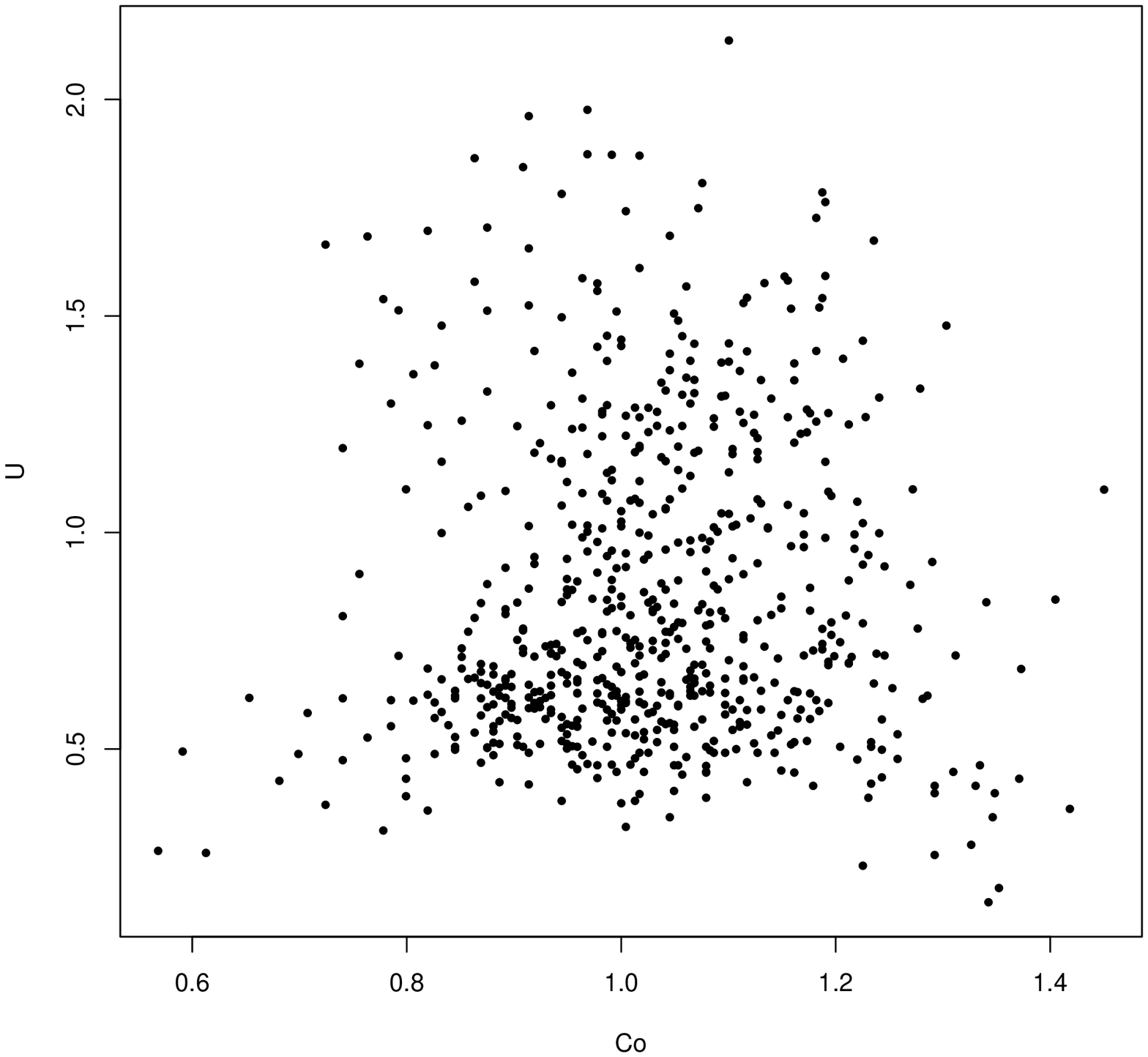}
\includegraphics[scale=0.3]{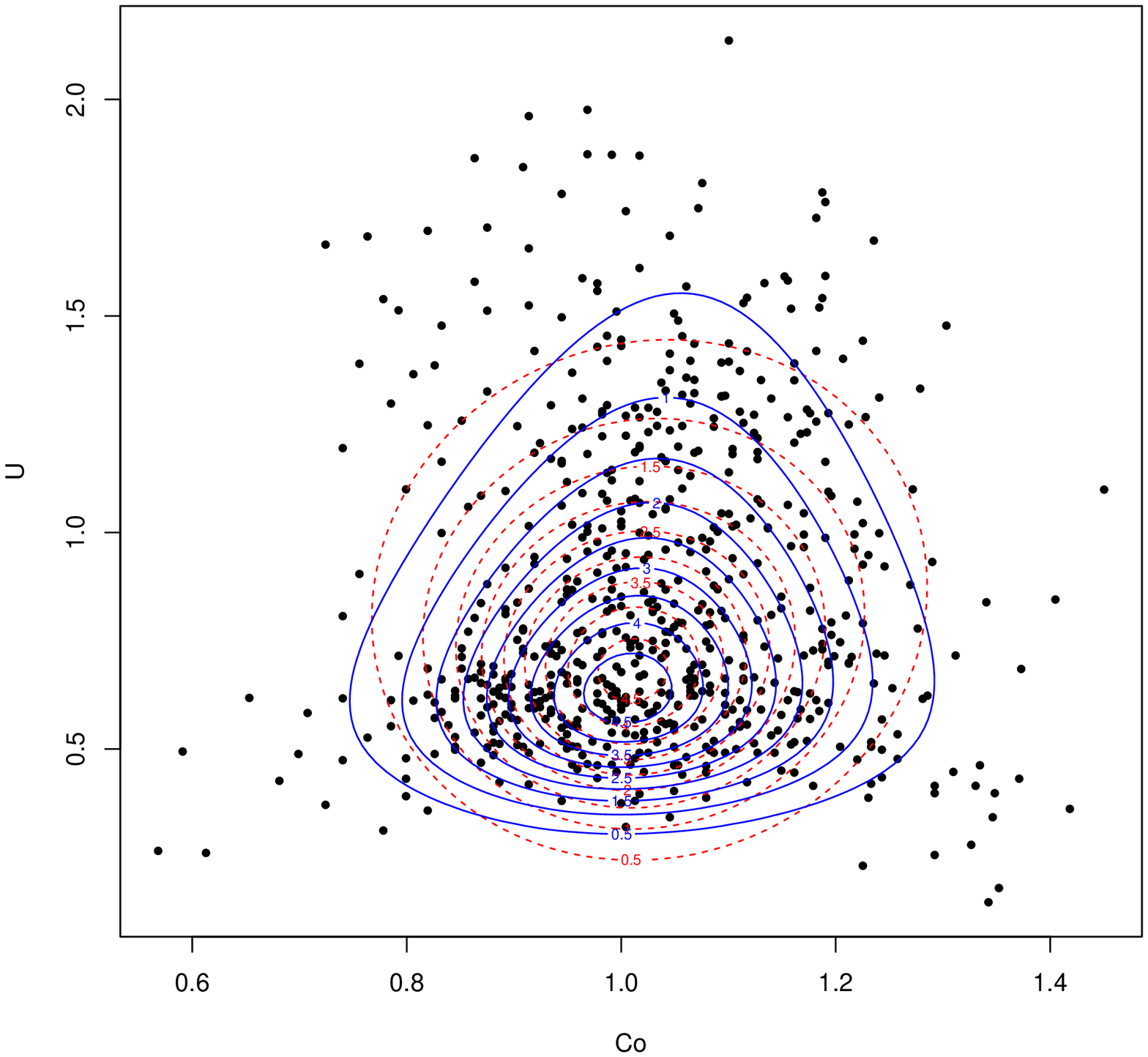} &
 \\
\end{tabular}
\caption{Scatterplot of petroleum data ($Co$ v. $U$). Right panel:  Comparison of standard NIG (red, dashed line) versus MSNIG (blue).
}
\label{fig:uran}
\end{figure}

This data consists of 655 petroleum samples collected from the
Montrose quadrangle of Western Colorado. The samples consist of
log-concentration readings for a number of chemical elements, and
are part of a multivariate dataset originally described by
\cite{cookjohnson81}.  The dataset is often used to compare and
contrast different copula approaches \citep{genestrivest93}.  For
ease of analysis and presentation we concentrate on two of the
elements Cobalt ($Co$) and Uranium ($U$).  Figure~\ref{fig:uran}
provides a scatterplot of the data overlaid with contour lines for
the standard NIG (red dashed) and multiple scaled NIG (blue)
displayed. From the contour lines we can see that the multiple
scaled NIG provides a better fit to the data and this is also
evidenced by significantly higher likelihood and BIC estimates for
the multiple scaled NIG ($\mathcal{L}$ = 207.5,
BIC  = -387) compared to the standard NIG
($\mathcal{L}$ = 168.4, BIC = -334).

\begin{table}[htp]
\centering
\caption{Estimated parameters for MSNIG and NIG on the Petroleum data \\ ($Co$ v. $U$)}
\begin{tabular}{c|cc}
\hline 
Parameters & MSNIG & NIG \\ 
\hline 
\vspace{-2mm} &  &  \\
$\mub$ & (0.96,0.35) & (0.99,0.46) \\ 
\vspace{-2mm} &  &  \\
$\betab$ & (2.73,13.57) & (2.10,5.25) \\ 
\vspace{-2mm} &  & • \\
$\Db$ & $\begin{pmatrix}
0.06 & -0.99 \\ 
0.99 & 0.06
\end{pmatrix}$  & - \\ 
\vspace{-2mm} &  & • \\ 
$\Ab$ & $\textup{diag}(1.08,0.93)$ & - \\ 
\vspace{-2mm} &  & • \\
$\Sigma$ & - & $\begin{pmatrix}
0.51 & -0.01 \\ 
-0.01 & 1.97
\end{pmatrix}$ \\ 
\vspace{-2mm} &  & • \\
$\gammab$ & (8.17,14.69) & 8.77 \\ 
\vspace{-2mm} &  & • \\
$\delta$ & 0.28 & 0.33 \\
\vspace{-2mm} &  & • \\ 
Log-like & 207.6 & 168.4 \\ 
\vspace{-2mm} &  &  \\
\hline 
\end{tabular} 
\end{table}


\subsection{Lymphoma data}

\begin{figure}[htpb]
 \centering
\begin{tabular}{cc}
\includegraphics[scale=0.3]{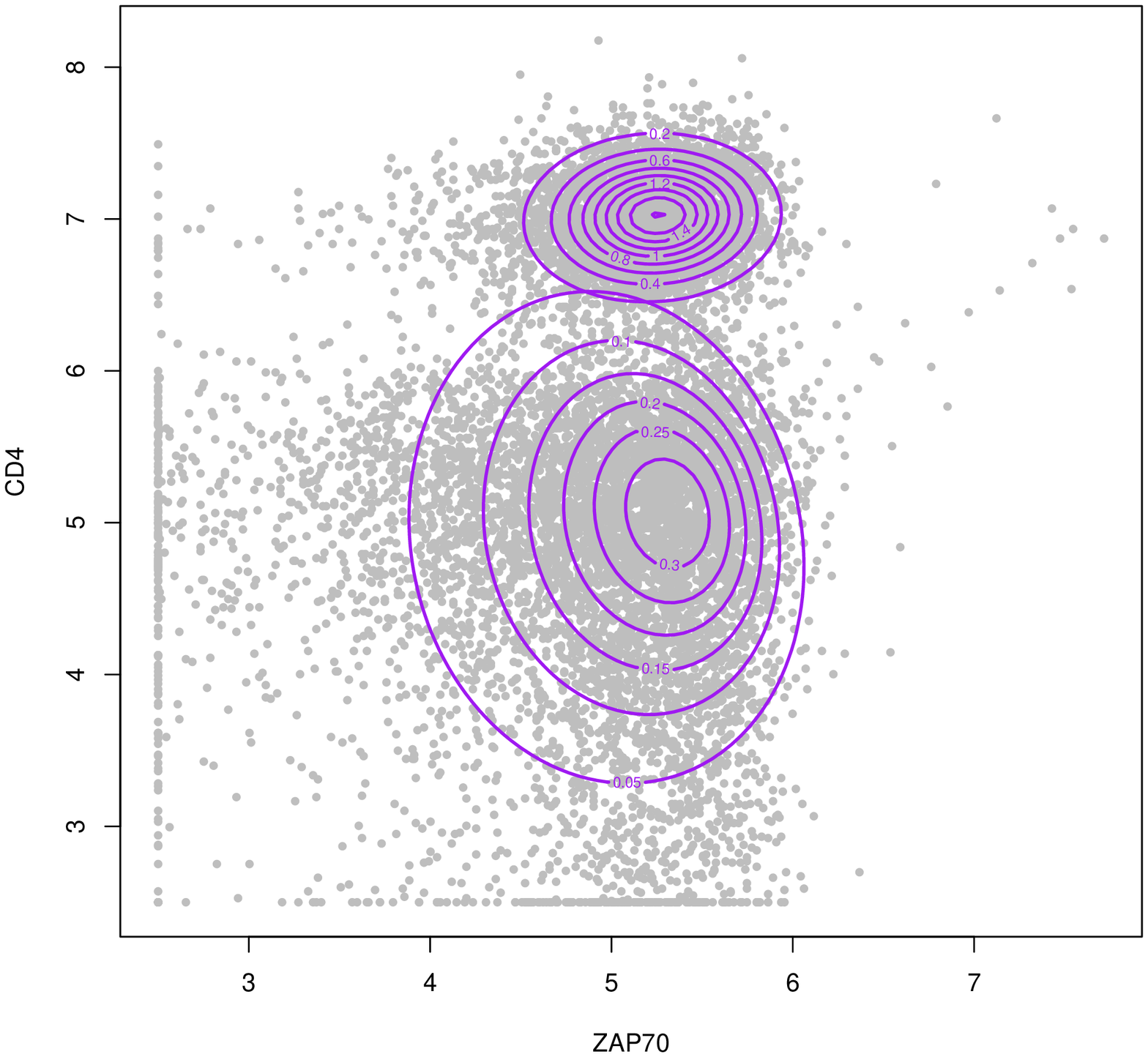} &  \hspace{-7mm}
\includegraphics[scale=0.3]{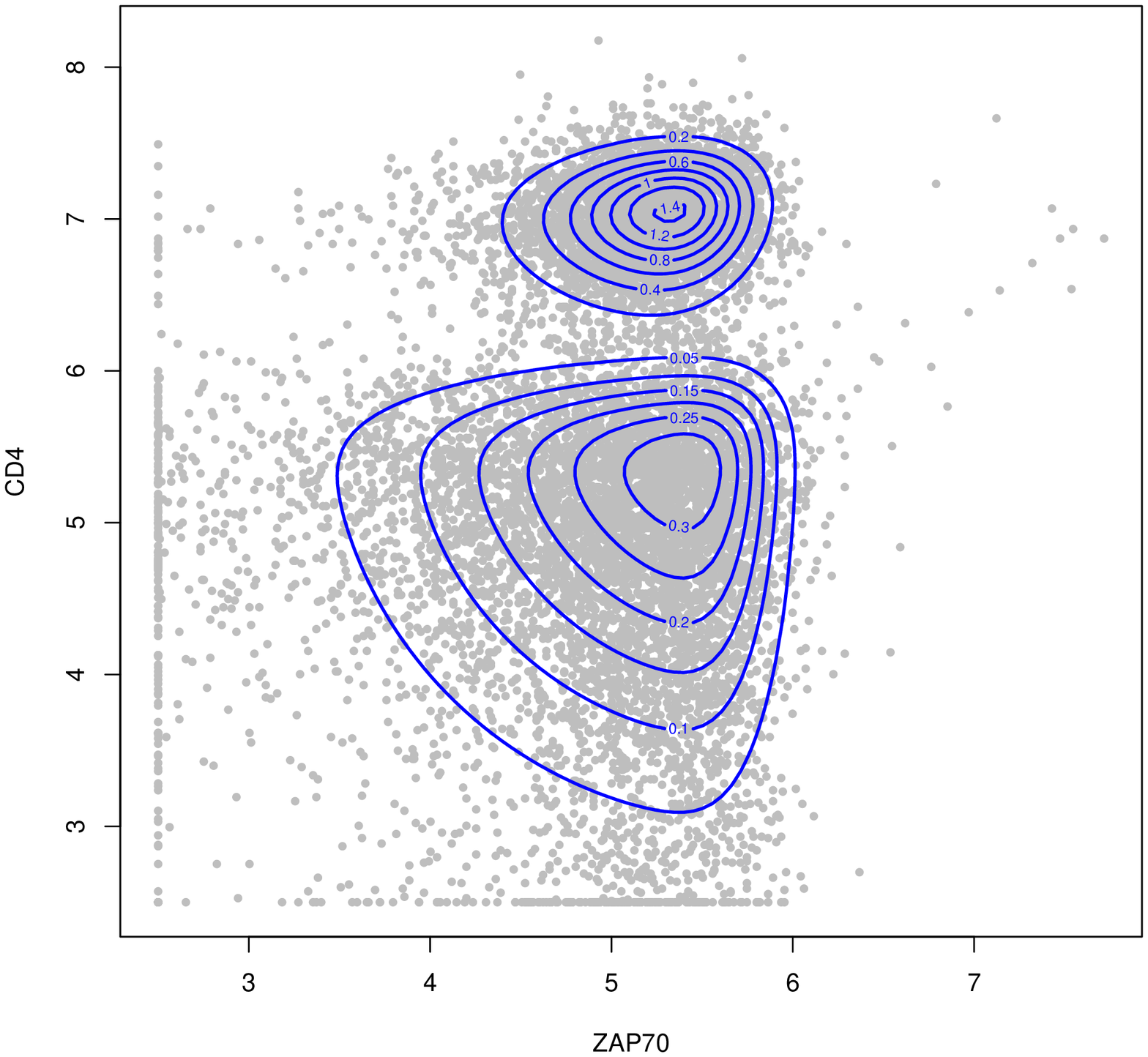} \\
(a) NIG & (b) {\it Unrestricted} Skew-$t$ \\
\includegraphics[scale=0.3]{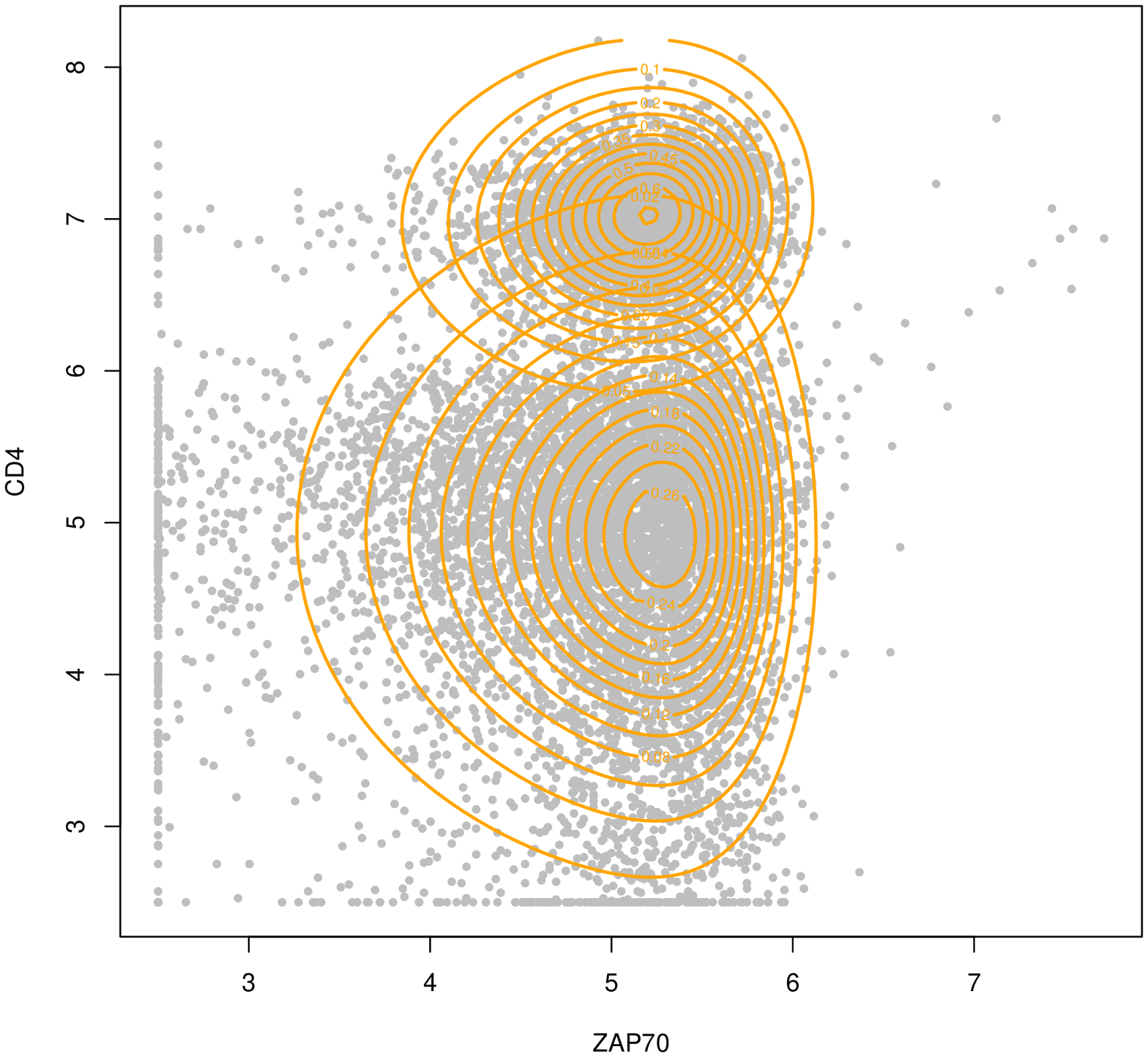} & \hspace{-7mm}
\includegraphics[scale=0.3]{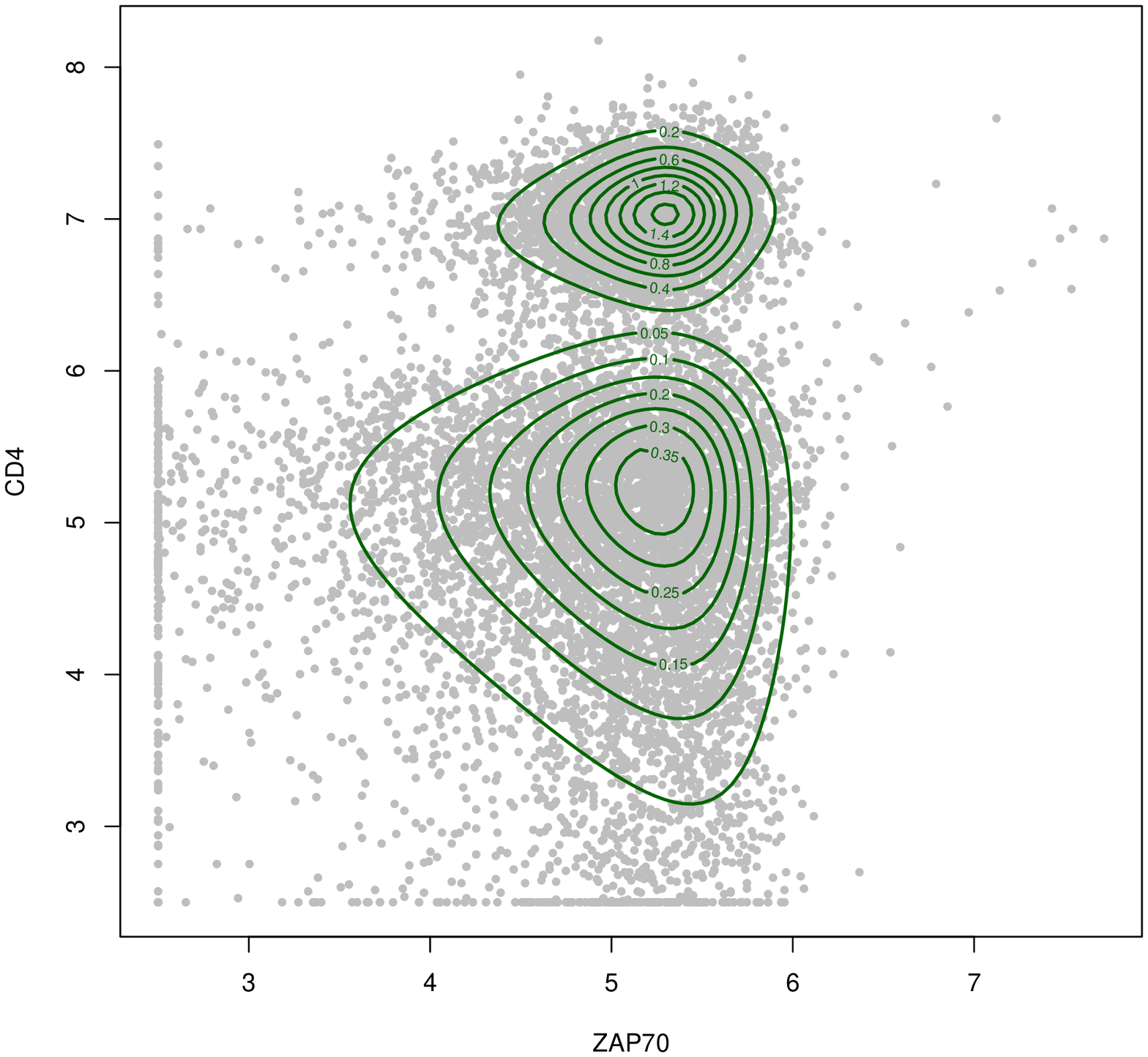} \\
(c) Skew-$t$ & (d) MSNIG \\
\end{tabular}
\caption{Lymphoma data, $CD4$ v. $ZAP70$. Fitted contour lines for: (a) Standard NIG \citep{KarlisSantourian2009}; (b) {\it Unrestricted} Skew-$t$ \citep{sahuetal03}; (c) Skew-$t$
\citep{azzalinietal96}; and (d) Multiple scaled NIG.
}
\label{fig:lymph}
\end{figure}

To further illustrate some of the differences between the standard
NIG and multiple scaled NIG we examine a clustering problem for a
lymphoma dataset recently analysed by \cite{leemclachlan12}. The
data consists of a subset of data originally presented and
collected by \cite{maieretal07}.  In \cite{maieretal07} blood
samples from 30 subjects were stained with four
fluorophore-labeled antibodies against $CD4$, $CD45RA$,
$SLP76(pY128)$, and $ZAP70(pY292)$ before and after an anti-$CD3$
stimulation. In the first example we will look at clustering a
subset of the data containing the variables $CD4$ and $ZAP70$
(Figure~\ref{fig:lymph}), which appear to be bimodal and
display an asymmetric pattern. In particular, one of the modes
appears to show both strong correlation between the two variables
and substantial skewness.

Of interest in this example is to compare the goodness of fit from
fitting mixtures of standard NIG and multiple scaled NIG
distributions.  For comparison, we also present the results of
fitting using mixtures of skew-normal \citep{lachosetal10} and
skew-$t$ distributions using two types of formulation: one in
which there is some separation between skewness and tail behaviour
and referred to as {\it unrestricted}
\citep{sahuetal03,leemclachlan12,lin10} and one with no such
separation \citep{azzalinietal96,bassoetal10,pyneetal09}.
Estimation of the parameters for these distributions was
undertaken using the R package \textbf{mixsmsn}
\citep{cabraleta2012} and for the unrestricted  skew-$t$ case
using R code available on:
\url{http://www.maths.uq.edu.au/~gjm/mix_soft/EMMIX-skew/index.html}\;
.

Figures~\ref{fig:lymph} (a) to (d) show the separate contour lines (of each component) from fitting mixtures of: standard NIG
\citep{KarlisSantourian2009}(a); unrestricted Skew-$t$ \citep{sahuetal03,leemclachlan12,lin10} (b); Skew-$t$ \citep{azzalinietal96,bassoetal10,pyneetal09} (c);
and multiple scaled NIG (d). Likelihood values and estimates of the BIC for the different approaches are also provided in Table~\ref{tab:reslymp1}.
As we can see from Figure~\ref{fig:lymph} there is quite a
difference in the goodness of fit between the approaches. In
particular, we see a clear difference in the fitted results
between the standard NIG and multiple scaled NIG with the latter
providing a closer fit to the data.  Similar results to the
standard NIG are obtained for the {\it no separation} Skew-$t$ (c)
and Skew-normal \citep{lachosetal10} (not shown) approaches.  Interestingly the fitted results
of the {\it unrestricted} Skew-$t$ (b) and the multiple scaled NIG
(d) appear to be similar. BIC values for these two approaches are
also similar (MSNIG = 47,175, {\it unrestr.} Skew-$t$ = 47,103) but with more support for the {\it unrestricted} Skew-$t$.

\begin{table}[htp]
\centering
\caption{Results for Lymphoma dataset} 
\begin{tabular}{l|cc|cc}
\hline 
 & \multicolumn{2}{c}{Example 1 ($CD4$ v. $ZAP70$)} & \multicolumn{2}{c}{Example 2 ($CD45$ v.$CD4$)}  \\
Model & Log-likelihood &  BIC & Log-likelihood &  BIC\\ 
\hline 
\vspace{-2mm} &   &   &  &  \\
MSNIG & -23,545 &  47,175 & -16,444 &  33,046 \\ 
\vspace{-3mm} &    & &    &  \\
NIG & -23,842 &  47,691 & -16,573 &  33,289\\
\vspace{-3mm} &    & &    &  \\ 
Skew-$t$ (Unrestr.) & -23,492 &  47,103 & -16,540 &  33,310 \\
\vspace{-3mm} &    &   &  &  \\ 
Skew-$t$ & -23,868 &  47,874 & -16,561 &  33,385 \\ 
\vspace{-3mm} &    & &    &  \\
Skew-normal & -23,762 & 47,663 & -16,573 &  33,410 \\ 
\vspace{-2mm} &    & &    &  \\
\hline 
\end{tabular}
\label{tab:reslymp1}
\end{table}

As noted by \cite{leemclachlan12} a possible reason for the
difference in the results between the unrestricted Skew-$t$ ({\it
e.g.} \cite{sahuetal03}) and the skew-normal and Skew-$t$ ({\it
e.g.} \citep{azzalinietal96}) is the differing degree of
dependency between the skewness parameter and the covariance for
the different approaches. As mentioned previously, in the skew-$t$
formulation of \cite{sahuetal03} there is some separation between
the skewness parameter and the covariance, which is not the case
for the other formulations of the skew-$t$ and skew-normal
approaches.

\begin{figure}[htpb]
 \centering
\begin{tabular}{cc}
\includegraphics[scale=0.30]{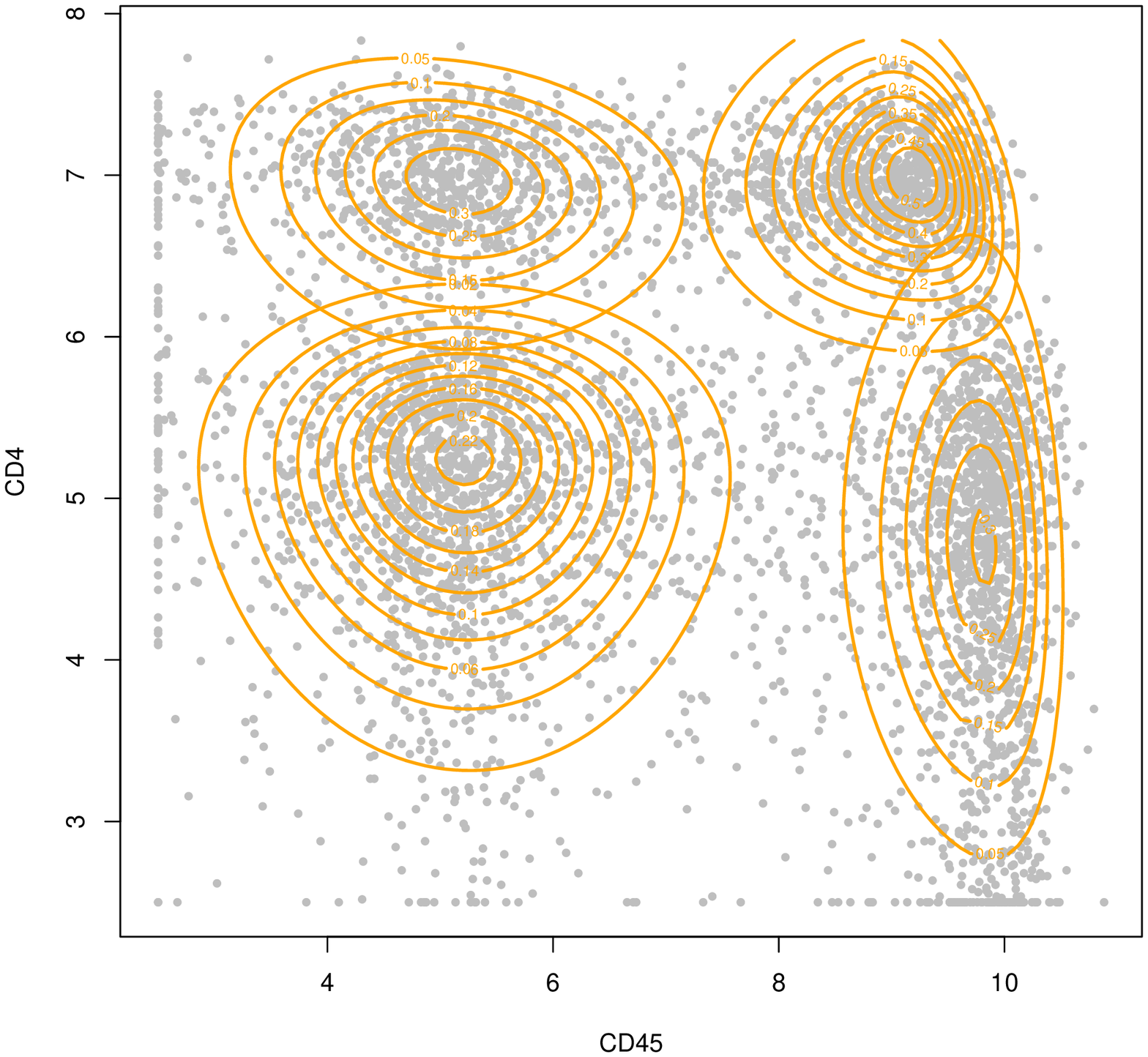} & \hspace{-7mm}
\includegraphics[scale=0.30]{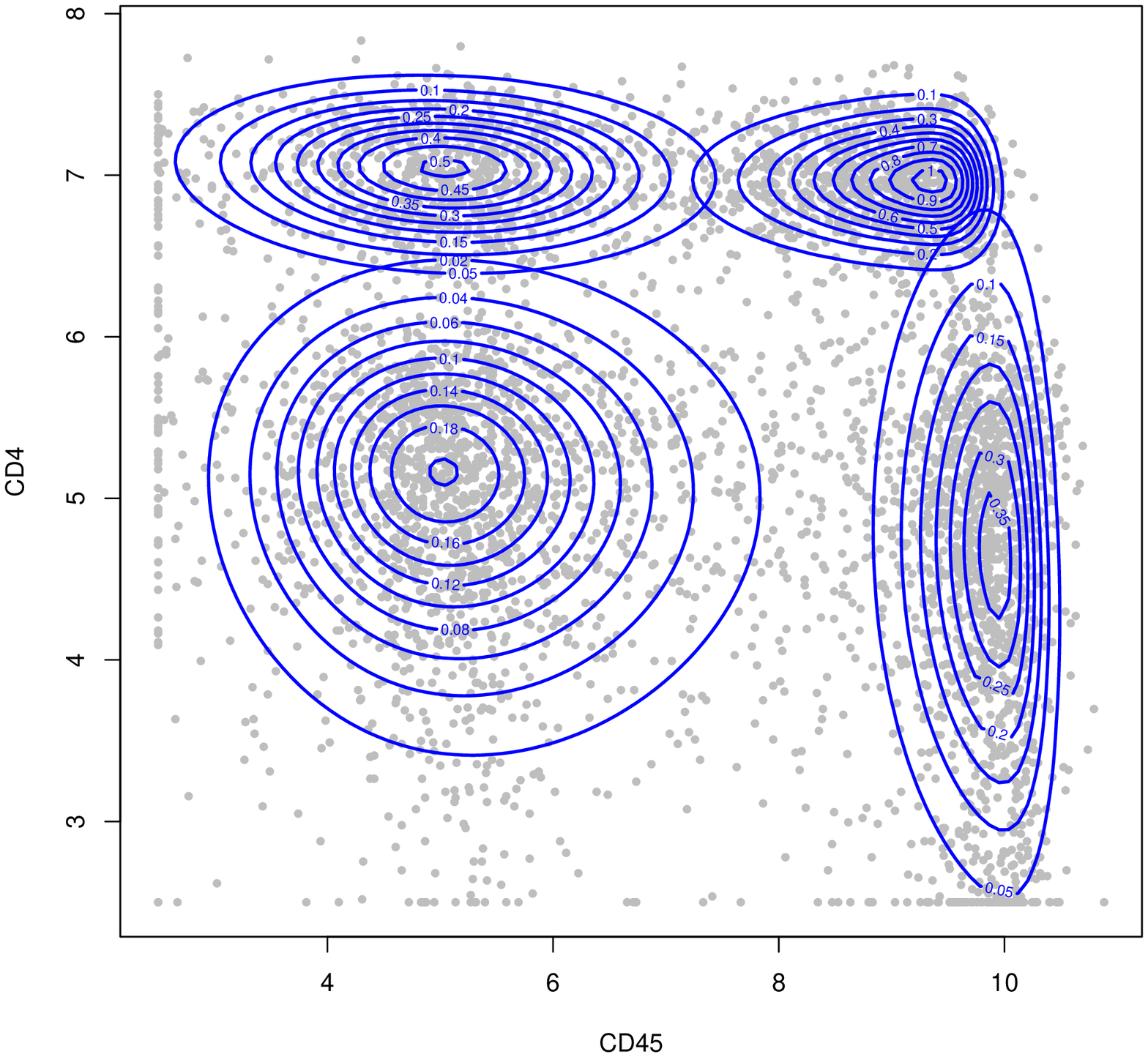} \\
(a) Skew-$t$ & (b) Skew-normal \\
\includegraphics[scale=0.30]{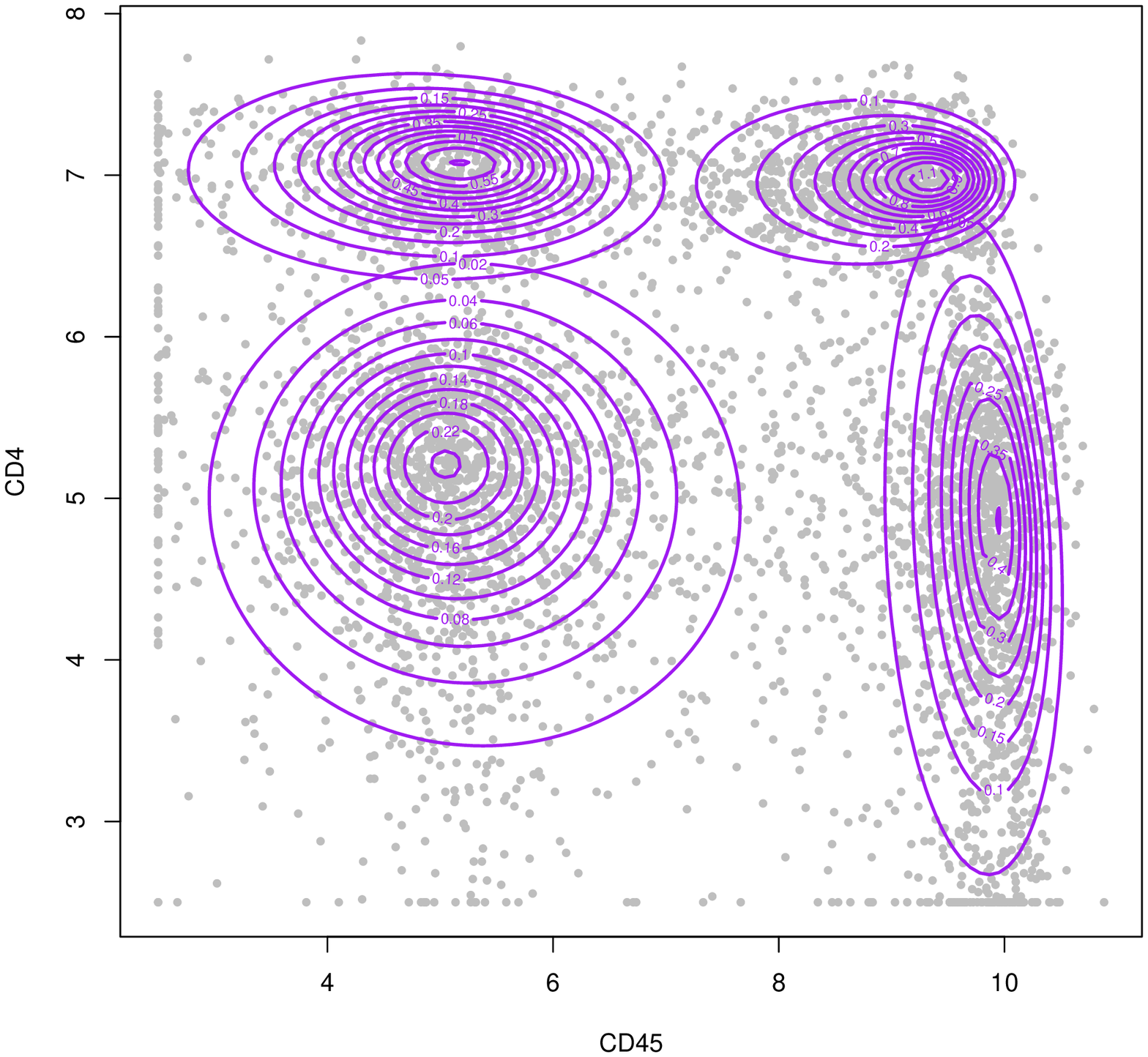} & \hspace{-7mm}
\includegraphics[scale=0.30]{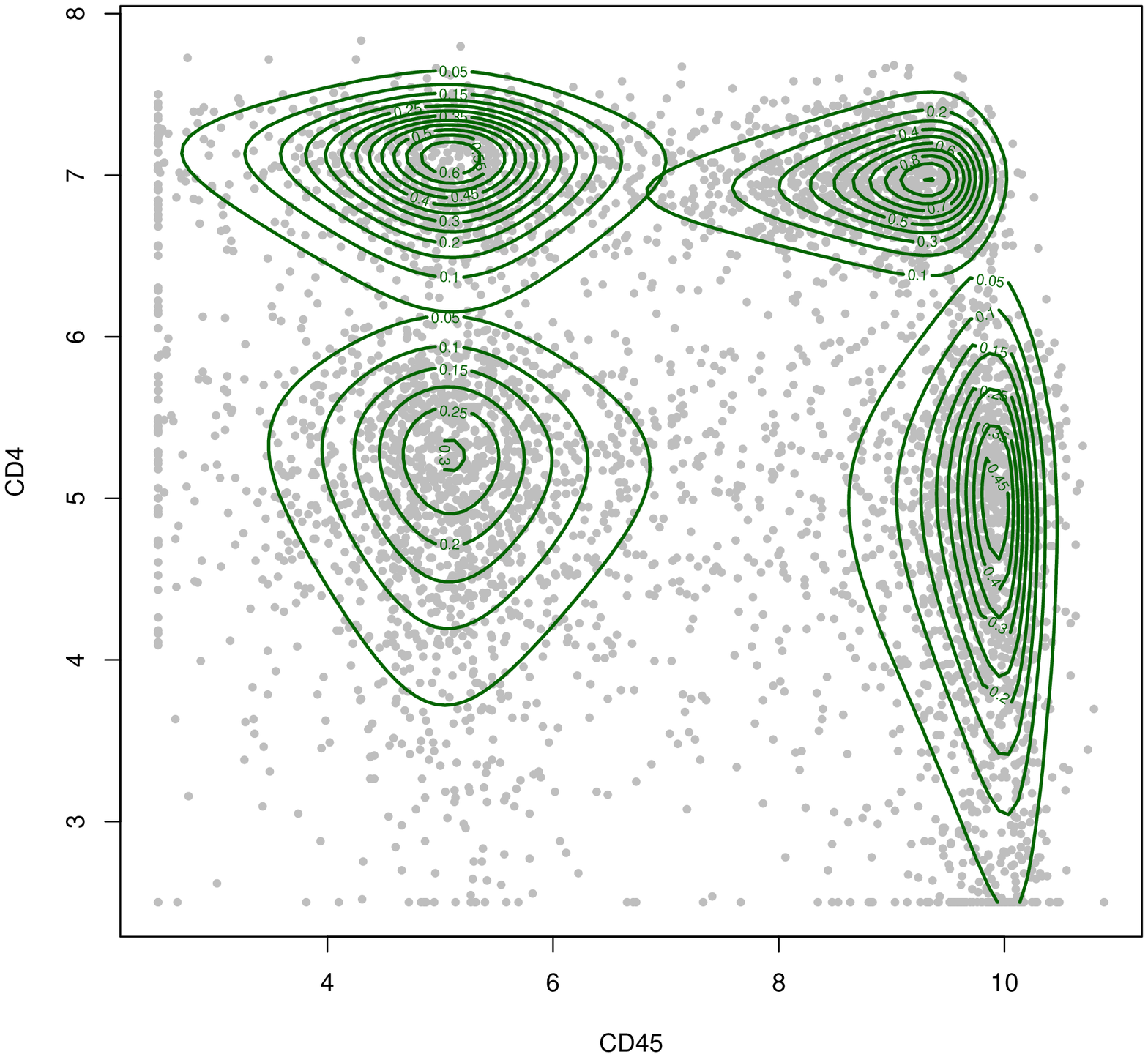} \\
(c) NIG & (d) MSNIG \\
\end{tabular}
\caption{Lymphoma data, $CD45$ v. $CD4$. Fitted contour lines for: (a) Skew-$t$;
 (b) {\it Unrestricted} Skew-$t$;  (c) Standard NIG; and (d) Multiple
scaled NIG.  } \label{fig:lymphalt}
\end{figure}

We now consider a second example to highlight further differences between the standard NIG and multiple scaled NIG in a clustering context using the same dataset.  In
this example we look at a subset of the dataset containing the
variables $CD45$ and $CD4$, which also appear to be highly
multimodal and asymmetric in shape.  The fitted results from a
mixture model with four components are shown in
Figure~\ref{fig:lymphalt} with contour lines representing the
fitted density of each component (see also results in Table~\ref{tab:reslymp1}).  From the fitted results we can
see a better fit from the multiple scaled NIG
(BIC = 33,046) compared to the
standard NIG (BIC = 33,289). The better fit appears
to come from the increased flexibility of the multiple scaled NIG
to represent non-elliptical shapes. The fitted results for the
Skew-$t$ and {\it unrestr.} Skew-$t$ do not appear to be better than for the standard NIG
(BIC = 33,385 and 33,310, respectively). Similar results to the Skew-$t$ are found
for the Skew-normal (not shown).

\section{Conclusion}\label{sec:disc}

We have proposed a relatively simple way to extend
\textit{location and scale mixture distributions}, such as the
multivariate generalised hyperbolic distribution (GH),  to allow
for different tail behaviour in each dimension. In contrast to
existing approaches, the approach has the advantage of: a closed
form density; allowing arbitrary correlation between dimensions;
and applicability to high dimensional spaces. Estimation of the
parameters of the multiple scaled GH (including the important
multiple scaled NIG as a particular case) is also relatively
straightforward using the familiar EM algorithm  and various
properties of the family are well defined. 
Assessments of the performance of the proposed model on simulated
and real data suggest that the extension provides a considerable
degree of freedom and flexibility in modelling data of varying
tail behaviour and directional shape.

For future research, parsimonious models could be considered using
special decompositions of the scale matrix such as in the
model-based clustering approach of \citet{CeleuxGovaert1995} and
\citet{FraleyRaftery2002}, 
which would be straightforward to generalize to
multiple scaled distributions. Similarly, for very high
dimensional data, other parsimonious models could also be
considered with a special modelling of the covariance matrix such
as in the High Dimensional Data Clustering (HDDC) framework of
\cite{Bouveyron2007}.

Although we have illustrated the approach on clustering examples,
the multiple scaled NIG is applicable to other contexts including, for example, regression modelling,
outlier detection and modelling of spatial data \citep{Forbes2010}.

\newpage

\section*{Appendices}

\subsection*{Appendix A: Multiple Scaled Normal Inverse Gaussian distribution (MSNIG)}\label{sec:mnig_mean}

In the case of the MSNIG where $W_m \sim {\cal
IG}(\lambda_m=-1/2,\gamma_m, \delta_m)$, expressions
(\ref{eqn:meanmgh}) and (\ref{eqn:varmgh}) simplify into

\begin{eqnarray}
E[\Yv_{MSNIG}] & = & \mub + \Db E[\Delta_W] \Ab \Db^{T} \betab \nonumber  \\
& = & \mub + \Db \text{diag}\bigg(\frac{\delta_1}{\gamma_1},\dots,\frac{\delta_M}{\gamma_M}\bigg) \Ab \Db^{T} \betab
\end{eqnarray}

\begin{align}
Var[\Yv_{MSNIG}]  = & \Db \text{diag}\bigg(\frac{\delta_1}{\gamma_1},\dots,\frac{\delta_M}{\gamma_M}\bigg) \Ab \Db^{T} \\ \notag
& +
\Db \text{diag}\bigg(\frac{\delta_1}{\gamma_1^3}[\Db^{T}\betab]_1^2,\dots,\frac{\delta_M}{\gamma_M^3}[\Db^{T}\betab]_M^2\bigg) \Ab \Db^{T} \nonumber  \\
 =& \Db \text{diag}\bigg(\frac{\delta_mA_m}{\gamma_m}\bigg)\bigg(1 + \frac{[\Db^{T}\betab]_m^2 \Ab_m}{\gamma_m^2}\bigg) \Db^{T}
\end{align}

\newpage

\subsection*{Appendix B: Algorithm for computing $\Db^{(r+1)}$}

\label{FG}

The goal is to minimize with respect to $\Db$ the following
quantity, where $\tilde{\Ab}$, $\mub$ and $\tilde{\betab}$ have
been fixed to current estimations namely $\tilde{\Ab}^{(r)}$,
$\mub^{(r+1)}$ and $\tilde{\betab}^{(r+1)}$,
\begin{align*}
\Db^{(r+1)} = & \arg\min_{\Db} f(\Db) \\
\mbox{where } \quad f(\Db) =&  \sum\limits_{i=1}^N
\text{trace}(\Db \Tb_{i}^{(r)}
\tilde{\Ab}^{-1(r)}\Db^{T}\Vb_{i})+\sum\limits_{i=1}^N
\text{trace}(\Db \Sb_{i}^{(r)} \tilde{\Ab}^{-1(r)}\Db^{T}\Bb_{i}) \\
& - 2(\sum\limits_{i=1}^N\text{trace}(\Db\tilde{\Ab}^{-1}\Db^{T}\Cb_{i})
\end{align*}
where $\Vb_{i}=(\yv_i- \mub^{(r+1)}) (\yv_i -\mub^{(r+1)})^T,
\Bb_{i}=\tilde{\betab}^{(r+1)} \tilde{\betab}^{T(r+1)}$,
$\Cb_{i}=(\yv_i- \mub^{(r+1)})\tilde{\betab}^{T(r+1)}$. Similarly
to \citet[see Appendix 2]{CeleuxGovaert1995}, we can derive from
\citet{FluryGautschi1986} the algorithm below. 

\vspace{.5cm} \noindent 
\textbf{Step 1.} We start from
an initial solution $\Db^0=[\db^0_1,\ldots,\db^0_M]$ where the
$\db^0_m$'s are $M$-dimensional orthonormal vectors.
\newline
 \textbf{Step 2.} For any couple $
(l,m)\in\{1,\ldots,M\}^2$ with $l\neq m$, the couple of vectors
$(\db_l,\db_m)$ is replaced with $(\deltab_l,\deltab_m)$ where
$\deltab_l=[\db_l,\db_m]\vb_1$ and $\deltab_m=[\db_l,\db_m]\vb_2$
with $\vb_1$ and $\vb_2$ two orthonormal vectors of $\Rset^2$ such
that $\vb_1$ is the eigenvector associated to the smallest
eigenvalue of the matrix
\begin{eqnarray*}
\Mb & = & \sum\limits_{i=1}^N
(\frac{t^{(r)}_{il}}{A_l^{(r)}}-\frac{t^{(r)}_{im}}{A_m^{(r)}})[d_l,d_m]^T\Vb_{i}[d_l,d_m]
+
\sum\limits_{i=1}^N(\frac{s^{(r)}_{il}}{A_l^{(r)}}-\frac{s^{(r)}_{im}}{A_m^{(r)}})[d_l,d_m]^T\Bb_{i}[d_l,d_m]  \\
& & - 2
\sum\limits_{i=1}^N(\frac{1}{A_l^{(r)}}-\frac{1}{A_m^{(r)}})[d_l,d_m]^T\Cb_{i}[d_l,d_m]
\end{eqnarray*}

\vspace{.3cm}

\noindent  {\bf Step 2} is repeated until it produces no decrease of the criterion $f(\Db)$.

\newpage

\subsection*{Appendix C: Mixture setting and estimation}


Denote the parameters of the mixture in the equivalent
parameterization (\ref{reprebis}) by $\phib=\{\pib,
\tilde{\psib}\}$ with $\tilde{\psib}=\{\tilde{\psib}_1, \ldots
\tilde{\psib}_K\}$ the mixture parameters with
$\tilde{\psib_k}=\{\mub_k, \Db_k,\tilde{\Ab}_k, \tilde{\betab}_k,
\tilde{\gammab}_k \}$ for $k=1 \ldots K$.
 For mixtures the EM algorithm
iterates over the following two steps.

\subsubsection*{E-step}

We denote by $\tau_{ik}^{(r)}$ the posterior probability that $\yv_i$ belongs to the $k$th component of the mixture given the current estimates of the mixture parameters $\phib^{(r)}$,

\begin{equation}
\tau^{(r)}_{ik}  = \frac{\pi^{(r)}_k
\mathcal{MSNIG}(\yv_i;\psib^{(r)}_k)}{p(\yv;\phib^{(r)})}
\end{equation}

The conditional expectation of the complete data log-likelihood $Q(\phib,\phib^{(r)})$ decomposes into three parts
\begin{equation}
Q(\phib,\phib^{(r)})=Q_{1}(\pib;\phib^{(r)})+Q_{2}(\tilde{\gammab};\phib^{(r)})+Q_{3}(\mub,\Db,\tilde{\Ab},\tilde{\betab},;\phib^{(r)})
\end{equation}
with
\begin{eqnarray}
Q_{1}(\pib;\phib^{(r)})  & = & \sum_{i=1}^{N}\sum_{k=1}^{K}\tau^{(r)}_{ik}\textup{log}\pi_k \\
Q_{2}(\tilde{\gammab};\phib^{(r)}) & =
&\sum\limits_{i=1}^N\sum\limits_{k=1}^K\tau^{(r)}_{ik}\sum\limits_{m=1}^M
E_{W_{im}}[\text{log}\mathcal{IG}(W_{im};\tilde{\gamma}_{km},1)|\yv_i,\phib^{(r)}]
\end{eqnarray}
and

\begin{align}
Q_{3}(\tilde{\gammab};\phib^{(r)})  = &
\sum\limits_{i=1}^N\sum\limits_{k=1}^K\tau^{(r)}_{ik}
E_{\Wb_{i}}[\text{log}\mathcal{N}_{M}(\mub_k + \\ \notag
& \Db_k
\Deltab_{\wv_i} \tilde{\Ab}_k \Db_k^T \tilde{\betab}_k, \Db_k
\Deltab_{\wv_i} \tilde{\Ab}_k \Db_k^T)|Z_{i}=k,\yv_i,\phib^{(r)}] \\
\nonumber  = &
\sum\limits_{i=1}^N\sum\limits_{k=1}^K\tau^{(r)}_{ik}
E_{\Wb_{i}}[-\frac{1}{2}(\yv_i-\mub_k-\Db_k\Deltab_{\wv_{i}}\Db^{T}_k\tilde{\betab}_k)^{T}\Db_k\tilde{\Ab}^{-1}\Deltab^{-1}_{\wv_{i}}\Db^{T}_k \\ \notag
& \times (\yv_i-
\mub_k-\Db_k\Deltab_{\wv_{i}}\Db^{T}_k\tilde{\betab}_k)|Z_{i}=k,\yv_i,\phib^{(r)} ] - \frac{1}{2}\text{log}|\tilde{\Ab}_k|  \nonumber
\end{align}
 ignoring constants.

Similarly to the E-step in Section \ref{estep}, the quantities
required for the E-step are given by,

\begin{eqnarray*}
s^{(r)}_{ikm} & = & E[W_{im}|Z_i=k,\yv_{i};\phib^{(r)})] =
\frac{\phi^{(r)}_{ikm}K_{0}(\phi^{(r)}_{ikm}\hat{\alpha}^{(r)}_{km})}{\hat{\alpha}^{(r)}_{km}K_{-1}(\phi^{(r)}_{ikm}\hat{\alpha}^{(r)}_{km})} \\
t^{(r)}_{ikm} & = & E[W^{-1}_{im}|Z_i=k,\yv_{i};\phib^{(r)})] =
\frac{\hat{\alpha}^{(r)}_{km}K_{-2}(\phi^{(r)}_{ikm}\hat{\alpha}^{(r)}_{km})}{\phi^{(r)}_{ikm}K_{-1}(\phi^{(r)}_{ikm}\hat{\alpha}^{(r)}_{km})}
\end{eqnarray*}
where
\begin{eqnarray*}
\phi_{ikm}^{(r)} & = & \displaystyle \sqrt{1+\frac{[\Db_k^{(r)T}(\yv_i-\mub_k^{(r)})]^{2}_{m}}{\tilde{\Ab}^{(r)}_{km}}} \\
\hat{\alpha}^{(r)}_{km} & = & \displaystyle
\sqrt{\tilde{\gammab}^{2(r)}_{km}+\frac{[\Db_k^{(r)T}\tilde{\betab}_k^{(r)}]^{2}_{m}}{\tilde{\Ab}^{(r)}_{km}}}
\end{eqnarray*}

\subsubsection*{M-step}

\textbf{Updating the $\pi_k$'s}.  The update of $\pib$ is
standard: for $k \in \{1,\dots,K\}, \hspace{2mm} \pi^{(r+1)}_k =
\displaystyle \frac{n_k}{N}$ where $n_k
=\sum\limits_{i=1}^{N}\tau^{(r)}_{ik}$.

\textbf{Updating the $\mub_k$'s}. It follows from the expression
of $Q_3$ that for $k \in \{1,\dots,K\}$, fixing $\Db_k$ to the
current estimation $\Db^{(r)}_k$, leads for all $m=1,\dots,M$ to
\begin{align*}
\mub^{(r+1)}_{km} =&
\bigg(\frac{\sum_{i=1}^{N}\tau_{ik}\Tb^{(r)}_{ik}\Db^{(r)T}_{k}}{n_k}-n_k
\;
(\sum_{i=1}^{N}\tau_{ik}\Sb^{(r)}_{ik})^{-1}\bigg)^{-1} \\ \notag
& \bigg(\frac{\sum_{i=1}^{N}
\tau_{ik}\Tb^{(r)}_{ik}\Db^{(r)T}_{k}\yv_i}{n_k}-\sum_{i=1}^{N}\tau_{ik}\yv_i\;
(\sum_{i=1}^{N}\tau_{ik}\Sb^{(r)}_{ik})^{-1}\bigg)
\end{align*}
where $\Tb^{(r)}_{ik} = \textup{diag}(t^{(r)}_{ik1} , . . . ,
t^{(r)}_{ikM})$ and $\Sb^{(r)}_{ik} = \textup{diag}(s^{(r)}
_{ik1},\dots,s^{(r)}_{ikM}$).

\textbf{Updating the $\tilde{\betab}_k$'s}. Similarly, it follows
from the expression of $Q_3$ that for $k \in \{1,\dots,K\}$,
fixing $\Db_k$ and $\mub_k$ to their current estimation
$\Db^{(r)}_k$ and $\mub^{(r)}_k$, leads to

\begin{eqnarray*}
\tilde{\betab}_k^{(r+1)}=\Db_k^{(r)}(\sum_{i=1}^{N}\tau_{ik}^{(r)}\Sb^{(r)}_{ik})^{-1}\Db^{(r)T}_k\sum_{i=1}^{N}\tau_{ik}^{(r)}(\yv_i-\mub_k^{(r+1)})
\end{eqnarray*}

\textbf{Updating the $\Db_k$'s}

The parameter $\Db_k$ is obtained by minimizing
\begin{align*}
\Db_k^{(r+1)}  = & \arg\min_{\Db_k} \bigg(\sum\limits_{i=1}^N
\text{trace}(\Db_k \Tb^{(r)}_{ik} \tilde{\Ab}_k^{(r)-1}
\Db_{k}^{T}\Vb_{ik})\\ \notag
& + \sum\limits_{i=1}^N \text{trace}(\Db_k \Sb^{(r)}_{ik} \tilde{\Ab}_k^{(r)-1}\Db_{k}^{T}\Bb_{ik}) - 2(\sum\limits_{i=1}^N\text{trace}(\Db_k\tilde{\Ab}_k^{(r)-1}\Db_{k}^{T}\Cb_{ik})\bigg)
\end{align*}

where $\Vb_{ik}=\tau^{(r)}_{ik}(\yv_i- \mub_k^{(r+1)}) (\yv_i
-\mub_k^{(r+1)})^T,
\Bb_{ik}=\tau^{(r)}_{ik}\tilde{\betab}^{(r+1)}_k
\tilde{\betab}^{(r+1)T}_k$ and
$\Cb_{ik}=\tau^{(r)}_{ik}(\yv_i-
\mub_k^{(r+1)})\tilde{\betab}^{(r+1)T}_k$


The parameter $\Db_k$ can be updated using an algorithm derived from Flury and Gautschi \citep[see][and Appendix B]{FluryGautschi1986}.

\noindent \textbf{Updating the $\tilde{\Ab}_k$'s.}  We have to
minimize the following quantity:

\begin{equation*}
\tilde{\Ab_k}^{(r+1)} = \arg\min_{\tilde{\Ab_k}}
(\text{trace}(\sum\limits_{i=1}^{N} \Mb_{ik}
\tilde{\Ab}_k^{-1})+\alpha_k \hspace{1mm}
\text{log}|\tilde{\Ab}_k|)
\end{equation*}
where 
$\Mb_{ik}=\Tb_{ik}^{(r)1/2}\Db_k^{(r+1)T}\Vb_{ik}\Db_k^{(r+1)}
\Tb_{ik}^{(r)1/2} + \Sb_{ik}^{(r)1/2}\Db_k^{(r+1)T}\Bb_{ik}
\Db_k^{(r+1)}
\Sb_{ik}^{(r)1/2}-\Db_k^{(r+1)T}(\Cb_{ik}+\Cb_{ik}^{T})\Db_k^{(r+1)}$
is a symmetric positive definite matrix and
$\alpha_k=\sum_{i=1}^{N}\tau_{ik}^{(r)}$

Using Corollary~\ref{cor:updateA} (see Section~\ref{sec:ML}) leads for all $m=1,\dots,M$ to

\begin{align*}
\tilde{\Ab}_{km}^{(r+1)}= & \frac{1}{\sum\limits_{i=1}^{N}\tau_{ik}^{(r)}}\sum\limits_{i=1}^{N}\tau_{ik}^{(r)}\bigg([\Db^{(r+1)T}_k(\yv_i-\mub_k^{(r+1)})]^{2}_m
t^{(r)}_{ikm}+[\Db^{(r+1)T}_k\tilde{\betab}_k^{(r+1)}]^{2}_m
s^{(r)}_{ikm} \\ \notag
& -2[\Db^{(r+1)T}_k(\yv_i-\mub_k^{(r+1)})]_m[\Db^{(r+1)T}_k\tilde{\betab}_k^{(r+1)}]_m\bigg)
\end{align*}

 \textbf{Updating the $\tilde{\gammab}_{k}$'s}. To
update $\tilde{\gammab}_{k}$ we have to minimize,

\begin{equation*}
\tilde{\gammab}_{k}^{(r+1)} = \arg\min_{\tilde{\gammab}} \bigg\{
\sum_{i=1}^{N}\tau_{ik}^{(r)}
\sum_{m=1}^{M}\frac{1}{2}\tilde{\gamma}_{km}^{2}s_{ikm}^{(r)} -
\tilde{\gamma}_{km})\bigg\} \label{eqn:mstep2ter}
\end{equation*}

which leads for all $m=1,\dots,M$ to

\begin{equation*}
\tilde{\gamma}^{(r+1)}_{km} = \displaystyle \frac{n_k}{\sum_{i=1}^{N} \tau_{ik} s_{ikm}}
\end{equation*}

To transform the estimated parameters back to the original ones, $
 \delta_k = |\tilde{\Ab}_k|^{\frac{1}{2M}},
 \gamma_{km}  = \tilde{\gammab}_{km}/\delta_k,
 \betab_k  =  \Db_k\tilde{\Ab}_k^{-1}\Db_k^{T}\tilde{\betab}_k,
 \Ab_k  = \tilde{\Ab}_k/|\tilde{\Ab}_k|^{\frac{1}{M}}$

\subsection*{Appendix D: Tail dependence}\label{sec:taildep}

Using \cite{colesetal99} and the R package `evd'
\citep{Rsoftware}, we assume that the data are {\it i.i.d.}
 random
vectors with common bivariate distribution function $G$, and we
define the  random vector $[X,Y]^T$ to be distributed according to
$G$.

The $\chi(q)$ plot is a plot of $q$  in  (0,1) (interpreted as a quantile level) against empirical estimates of function
\begin{equation}
\chi(q) = 2 -   − \textup{log}(p(F_X (X ) < q, F_Y (Y ) < q))/
\textup{log}(q)
\end{equation}
where $F_X$ and $F_Y$ are the marginal distribution functions. The
quantity $\chi(q)$ is bounded by
$$
2 - \textup{log}(2q - 1)/ \textup{log}(q) \leq \chi(q) \leq 1
$$
where the lower bound is interpreted as $-\infty$ for $q \leq 1/2$ and zero for $q = 1$.

The function $\chi(q)$ can be interpreted as a quantile dependent measure of dependence. In particular,
the sign of $\chi(q)$ determines whether the variables are positively or negatively associated at quantile
level $q$.


\bibliographystyle{elsarticle-harv} 
\bibliography{multiscaledistbib,skeweddist}

\end{document}